\documentclass[aps,prd,twocolumn,superscriptaddress,showpacs]{revtex4}
\usepackage[dvips]{graphicx}
\usepackage{amsmath,amssymb,times}

\newcommand{\bequ}{\begin{equation}}
\newcommand{\eequ}{\end{equation}}
\newcommand{\bea}{\begin{eqnarray}}
\newcommand{\eea}{\end{eqnarray}}

\DeclareSymbolFont{boldletters}{OML}{cmm} {b}{it}
\DeclareSymbolFontAlphabet{\mathbit}{boldletters}
\DeclareMathSymbol{\alpha}{\mathalpha}{letters}{"0B}
\DeclareMathSymbol{\beta}{\mathalpha}{letters}{"0C}
\DeclareMathSymbol{\gamma}{\mathalpha}{letters}{"0D}
\DeclareMathSymbol{\delta}{\mathalpha}{letters}{"0E}
\DeclareMathSymbol{\epsilon}{\mathalpha}{letters}{"0F}
\DeclareMathSymbol{\zeta}{\mathalpha}{letters}{"10}
\DeclareMathSymbol{\eta}{\mathalpha}{letters}{"11}
\DeclareMathSymbol{\theta}{\mathalpha}{letters}{"12}
\DeclareMathSymbol{\iota}{\mathalpha}{letters}{"13}
\DeclareMathSymbol{\kappa}{\mathalpha}{letters}{"14}
\DeclareMathSymbol{\lambda}{\mathalpha}{letters}{"15}
\DeclareMathSymbol{\mu}{\mathalpha}{letters}{"16}
\DeclareMathSymbol{\nu}{\mathalpha}{letters}{"17}
\DeclareMathSymbol{\xi}{\mathalpha}{letters}{"18}
\DeclareMathSymbol{\pi}{\mathalpha}{letters}{"19}
\DeclareMathSymbol{\rho}{\mathalpha}{letters}{"1A}
\DeclareMathSymbol{\sigma}{\mathalpha}{letters}{"1B}
\DeclareMathSymbol{\tau}{\mathalpha}{letters}{"1C}
\DeclareMathSymbol{\upsilon}{\mathalpha}{letters}{"1D}
\DeclareMathSymbol{\phi}{\mathalpha}{letters}{"1E}
\DeclareMathSymbol{\chi}{\mathalpha}{letters}{"1F}
\DeclareMathSymbol{\psi}{\mathalpha}{letters}{"20}
\DeclareMathSymbol{\omega}{\mathalpha}{letters}{"21}
\DeclareMathSymbol{\varepsilon}{\mathalpha}{letters}{"22}
\DeclareMathSymbol{\vartheta}{\mathalpha}{letters}{"23}
\DeclareMathSymbol{\varpi}{\mathalpha}{letters}{"24}
\DeclareMathSymbol{\varrho}{\mathalpha}{letters}{"25}
\DeclareMathSymbol{\varsigma}{\mathalpha}{letters}{"26}
\DeclareMathSymbol{\varphi}{\mathalpha}{letters}{"27}
\DeclareMathSymbol{\Gamma}{\mathalpha}{letters}{"00}
\DeclareMathSymbol{\Delta}{\mathalpha}{letters}{"01}
\DeclareMathSymbol{\Theta}{\mathalpha}{letters}{"02}
\DeclareMathSymbol{\Lambda}{\mathalpha}{letters}{"03}
\DeclareMathSymbol{\Xi}{\mathalpha}{letters}{"04}
\DeclareMathSymbol{\Pi}{\mathalpha}{letters}{"05}
\DeclareMathSymbol{\Sigma}{\mathalpha}{letters}{"06}
\DeclareMathSymbol{\Upsilon}{\mathalpha}{letters}{"07}
\DeclareMathSymbol{\Phi}{\mathalpha}{letters}{"08}
\DeclareMathSymbol{\Psi}{\mathalpha}{letters}{"09}
\DeclareMathSymbol{\Omega}{\mathalpha}{letters}{"0A}

\begin{document}
\preprint{SAGA-HE-244-08}
\title{Vector-type four-quark interaction and its impact on QCD phase structure}

\author{Yuji Sakai}
\email[]{sakai@phys.kyushu-u.ac.jp}
\affiliation{Department of Physics, Graduate School of Sciences, Kyushu University,
             Fukuoka 812-8581, Japan}
\author{Kouji Kashiwa}
\email[]{kashiwa@phys.kyushu-u.ac.jp}
\affiliation{Department of Physics, Graduate School of Sciences, Kyushu University,
             Fukuoka 812-8581, Japan}

\author{Hiroaki Kouno}
\email[]{kounoh@cc.saga-u.ac.jp}
\affiliation{Department of Physics, Saga University,
             Saga 840-8502, Japan}

\author{Masayuki Matsuzaki}
\email[]{matsuza@fukuoka-edu.ac.jp}
\affiliation{Department of Physics, Fukuoka University of Education, 
             Munakata, Fukuoka 811-4192, Japan}

\author{Masanobu Yahiro}
\email[]{yahiro@phys.kyushu-u.ac.jp}
\affiliation{Department of Physics, Graduate School of Sciences, Kyushu University,
             Fukuoka 812-8581, Japan}

\date{\today}

\begin{abstract}
Effects of the vector-type four-quark interaction on QCD phase structure 
are investigated in the imaginary chemical potential ($\mu$) region, 
by using 
the Polyakov-loop extended Nambu--Jona-Lasinio (PNJL) model with 
the extended ${\mathbb Z}_{3}$ symmetry. 
In the course to this end, we clarify analytically the Roberge-Weiss periodicity 
and symmetry properties of various quantities under the existence of a vector-type 
four-quark interaction. 
In the imaginary $\mu$ region, the chiral condensate and 
the quark number density are sensitive to the strength of the interaction. 
Based on this result, we propose a possibility to determine 
the strength of the vector-type interaction, which largely affects QCD phase structure 
in the real $\mu$ region, by comparing the results of lattice simulations and 
effective model calculations in the imaginary $\mu$ region. 
\end{abstract}

\maketitle


\section{Introduction}
\label{Introduction}
The progress in computer power has made it feasible to do realistic simulations in the lattice QCD for finite temperature ($T$) system 
without quark chemical potential ($\mu$)~\cite{Kog}. 
As for $\mu^2>0$, however, 
lattice QCD has the well-known sign problem, and then 
the results are still far from perfection; for example, see 
Ref.~\cite{Kogut2} and references therein. 
Several approaches have been proposed to solve the sign problem. 
One of them is the use of imaginary chemical potential, 
since the fermionic determinant that appears in the Euclidean partition 
function is real in the case; for example, see 
Refs.~\cite{FP,Elia,Chen} and references therein.

When physical quantities are available with lattice QCD 
in the imaginary $\mu$ region, 
in principle it is possible  to extrapolate them to real $\mu$, 
until there appears a discontinuity. Actually, such an  
extrapolation was made for the phase transition curve 
by assuming some analytic functions for the curve 
\cite{FP,Elia}. 
This direct extrapolation may work for small 
real $\mu/T$, but its accuracy is quite unknown 
for large real $\mu/T$ \cite{Sakai}. This problem may be circumvented by 
the effective theory that can evaluate the partition function 
in both the real and imaginary $\mu$ regions and 
reproduce the results of lattice QCD in the imaginary $\mu$ region, 
if such an effective theory is found.

In the region of imaginary chemical potential $\mu=iT\theta$, 
Roberge and Weiss (RW) found \cite{RW} that the partition function 
$Z(\theta)$ of SU($N$) gauge theory is a periodic function of $\theta$ 
with a period $2\pi/N$, 
showing that $Z(\theta+2{\pi}k/N)$ is reduced to $Z(\theta)$ with 
the ${\mathbb Z}_{N}$ transformation for any integer $k$. 
The RW periodicity means that 
$Z(\theta)$ is invariant under 
the combination of 
the ${\mathbb Z}_{N}$ transformation and another transformation 
$\theta \to \theta +  2 \pi k/N$, that is, under 
the extended ${\mathbb Z}_{N}$ transformation \cite{Sakai}
\bea
q \to U q, ~
A_{\nu} \to UA_{\nu}U^{-1} - \tfrac{i}{g} (\partial_{\nu}U)U^{-1}, ~
\theta \to \theta +  \tfrac{2 \pi k}{N}, 
\label{extended-z3}
\eea
where $U(x,\tau)$ are elements of SU($N$) with the boundary condition 
$
U(x,\beta)=\exp(-2i \pi k/N)U(x,0) , 
$
$q$ is the fermion field, $A_\nu$ is the gauge field and 
$\beta$ is the inverse of temperature $T$.  
Quantities invariant under the extended ${\mathbb Z}_N$ 
transformation, such as the thermodynamic potential $\Omega(\theta)$ and 
the chiral condensate, keep the RW periodicity. 
Meanwhile, the Polyakov loop $\Phi$ is not invariant under 
the transformation (\ref{extended-z3}), since 
it is transformed as 
$\Phi \to \Phi \exp(-i2\pi k/N)$. 
Such a non-invariant quantity does not have the periodicity, but 
this problem can be solved 
by the modified Polyakov loop defined later in Eq. (\ref{eq:NE11}) 
invariant under the extended ${\mathbb Z}_N$ transformation \cite{Sakai}. 
Roberge and Weiss also showed with perturbation that in the high $T$ region 
$d\Omega(\theta)/d\theta$ and $\Phi(\theta)$ are discontinuous at 
$\theta={(2k+1)\pi/N}$, and 
also found with the strongly coupled lattice theory that 
the discontinuities disappear in the low $T$ region. 
The first-order RW phase transition of 
$d\Omega(\theta)/d\theta$, that is, the quark number density 
$n=-d\Omega/d\mu=i(d\Omega(\theta)/d\theta)/T$ 
was observed in lattice simulations~\cite{FP,Elia,Chen}.

The Nambu--Jona-Lasinio (NJL) model~\cite{NJ1} is famous 
as a model for understanding chiral symmetry breaking, 
but it does not have the extended ${\mathbb Z}_3$ symmetry 
as well as the confinement mechanism. 
In the previous work \cite{Sakai}, we pointed out that 
the Polyakov-loop extended NJL (PNJL) model designed to 
have the confinement mechanism 
\cite{Meisinger,Dumitru,Fukushima,Ghos,Megias,
Ratti1,Ciminale,Ratti2,Rossner,Hansen,Sasaki1,Schaefer,Fu,Kashiwa3,Costa,Fukushima2,Kashiwa4,Abuki}
possesses the extended ${\mathbb Z}_3$ symmetry. 
Hence the thermodynamic potential $\Omega(\theta)$ of the PNJL model 
has the RW periodicity. Using the RW periodicity and the $\theta$-evenness of 
$\Omega(\theta)$, $\Omega(\theta)=\Omega(-\theta)$, we also found that 
$\theta$-odd ($\theta$-even) quantities such as $n$ ($\sigma$) 
have first-order (second-order) phase transitions 
on lines $\theta=(2k+1)\pi/3$ in the $\theta$-$T$ plane. 
This result is consistent with the results of lattice QCD and even more 
informative. Thus, the extended ${\mathbb Z}_3$ symmetry is essential. 
The extended ${\mathbb Z}_3$ symmetry of finite $\theta$ QCD is 
an extension of the ${\mathbb Z}_3$ symmetry of pure gauge QCD.

The confinement mechanism is a key to understand the physical 
real $\mu$ world. 
Actually, it is predicted that the mechanism largely shifts 
the critical endpoint~\cite{AY} toward  higher $T$ and lower $\mu$ 
than the NJL model predicts~\cite{Rossner,Kashiwa3,Fukushima2}. 
In contrast, it is known \cite{KKKN,Kashiwa,Kashiwa3,Fukushima2} that 
the vector-type four-quark interaction $(\bar{q}\gamma_\mu q)^2$ largely moves 
the critical endpoint in the opposite direction, if it is newly added 
to the NJL and PNJL models. Thus, it is essential 
from the phenomenological point of view to determine the strength 
of the vector-type four-quark interaction. 

So far, the vector-type interaction was often ignored 
in the NJL and PNJL models. 
In the relativistic meson-nucleon theory~\cite{Walecka}, meanwhile, 
the repulsive force mediated by vector mesons 
is essential to account for the saturation property of nuclear matter. 
Using the auxiliary field method, one 
can convert quark-quark interactions 
to meson-quark interactions; 
for example, see Refs.~\cite{KS,Sakaetal,KSKHTMY} and references therein. 
In the hadron phase, furthermore, quarks have a large effective mass as a result 
of spontaneous chiral symmetry breaking, and then nucleons can be 
considered to be formed from such three heavy quarks, that is, 
three constituent quarks. 
It is then natural to think that there exists 
a correspondence between the meson-nucleon interactions and 
the quark-quark interactions. 
In this point of view, it is quite likely that 
the vector-type four-quark interaction 
is not negligible and 
even significant in particular 
at the finite quark-density region corresponding 
to the nuclear saturation density. 

In this paper we investigate roles of the vector-type four-quark interaction 
in the imaginary $\mu$ region and show that the chiral condensate $\sigma$ and 
the quark number density $n$ are 
quantities sensitive to the strength of 
vector-type four-quark interaction. 
We then propose to measure $n$ as well as $\sigma$ with lattice QCD. 
If the strength of vector-type four-quark interaction is determined 
from the measured $\sigma$ and $n$, the PNJL model with the interaction can predict 
the phase diagram in the real $\mu$ region with reasonable reliability. 
The previous proof \cite{Sakai}  on the RW periodicity 
and the even/odd property of 
the extended ${\mathbb Z}_{3}$ invariant quantities  
is not applicable to the case with vector-type four-quark interaction. 
We then prove the properties in a way different 
from the previous one \cite{Sakai}. 

In sec. \ref{PNJL}, we present the PNJL model with 
the vector-type four-quark interaction, and derive the 
thermodynamic potential invariant under 
the extended ${\mathbb Z}_{3}$ transformation. 
For the case of imaginary $\mu=iT\theta$, it is proven that 
the extended ${\mathbb Z}_{3}$ invariant quantities
exhibit the RW periodicity. For both of imaginary and real $\mu$ cases, 
their even/odd properties are determined. 
It is shown from the the RW periodicity and the even/odd properties 
that the RW phase transition appearing in $\theta=(2k+1)\pi/3$ 
is a family of first-order phase transitions in $\theta$-odd 
quantities and second-order ones in $\theta$-even quantities. 
In sec. \ref{Numerical-results}, we show results of numerical calculations.
Section \ref{Summary} is devoted to Summary.

\section{PNJL model}
\label{PNJL}
\subsection{Formulation}
\label{Formulation}

We consider the two-flavor PNJL model with 
the vector-type four-quark interaction. 
The PNJL Lagrangian is 
\begin{align}
 {\cal L}  =& {\bar q}(i \gamma_\nu D^\nu -m_0)q 
  + G_{\rm s}[({\bar q}q)^2 +({\bar q}i\gamma_5 {\vec \tau}q)^2] \notag\\
             &\hspace{3mm}  -G_{\rm v}({\bar q}\gamma_\mu q)^2
              - {\cal U}(\Phi [A],{\Phi_{\rm H}} [A],T) ,
             \label{eq:E1}
\end{align}
where $q$ denotes the quark field, 
$m_0$ does the current quark mass, 
${\vec \tau}$ stands for the isospin matrix, and 
$D^\nu=\partial^\nu-iA^\nu - \delta_{0}^{\nu}\mu$ 
for the complex chemical potential $\mu=\mu_{\rm R}+iT\theta$. 
Parameters $G_{\rm s}$ and $G_{\rm v}$ represent the coupling constants 
of the scalar- and vector-type four-quark interactions, 
$({\bar q}q)^2$ and $({\bar q}\gamma_\mu q)^2$, respectively. 
The Polyakov potential ${\cal U}$, defined later in ~\eqref{eq:E13} , 
is a function of the Polyakov loop $\Phi$ 
and its Hermitian conjugate $\Phi_{\rm H}$, 
\begin{align}
\Phi      = {1\over{N}}{\rm Tr} L,~~~~
\Phi_{\rm H}  = {1\over{N}} {\rm Tr}L^\dag ,
\end{align}
with
\begin{align}
L({\bf x})  = {\cal P} \exp\Bigl[
                {i\int^\beta_0 d \tau A_4({\bf x},\tau)}\Bigr] ,
\end{align}
where $A_4 = iA_0$, $N=3$, and ${\cal P}$ is the path ordering. 
In the PNJL model, $\Phi$ and $\Phi_{\rm H}$ are 
treated as classical variables. 
The temporal component $A_4$ is diagonal 
in the flavor space, because the color and the flavor space 
are completely separated out in the present case. 
In the Polyakov gauge, $L$ can be written in a diagonal form 
in the color space~\cite{Fukushima}: 
\begin{align}
L 
=  e^{i \beta (\phi_3 \lambda_3 + \phi_8 \lambda_8)}
= {\rm diag} (e^{i \beta \phi_a},e^{i \beta \phi_b},
e^{i \beta \phi_c} ),
\label{eq:E6}
\end{align}
where $\phi_a=\phi_3+\phi_8/\sqrt{3}$, $\phi_b=-\phi_3+\phi_8/\sqrt{3}$
and $\phi_c=-(\phi_a+\phi_b)=-2\phi_8/\sqrt{3}$.

The Polyakov loop $\Phi$ is an exact order parameter of the spontaneous 
${\mathbb Z}_3$ symmetry breaking in the pure gauge theory.
Although the ${\mathbb Z}_3$ symmetry is not exact
in the system with dynamical quarks, it still seems to be a good indicator of 
the deconfinement phase transition. 
Therefore, we use $\Phi$ to define the deconfinement phase transition.

Making the mean field approximation (MFA), one can get 
the Lagrangian density 
\begin{align}
{\cal L}_{\rm MFA} =& 
{\bar q}( i \gamma_\mu D^\mu+\gamma_0\Sigma_{\rm v}- (m_0+\Sigma_{\rm s}) )q \notag\\
&\hspace{10mm} - U(\sigma ,n) - {\cal U}(\Phi,\Phi_{\rm H},T) , 
\label{eq:E7} 
\end{align}
with 
\begin{eqnarray}
\sigma &=& \langle \bar{q}q \rangle, ~~~
n=\langle {\bar q}\gamma_0 q \rangle, ~~~
\Sigma_{\rm s}=-2G_{\rm s}\sigma,
\notag\\
\Sigma_{\rm v} &=& -2G_{\rm v}n,~~~
U= G_{\rm s} \sigma^2-G_{\rm v}n^2 , 
\label{eq:E10}
\end{eqnarray}
and obtain 
the thermodynamic potential $\Omega$ 
by making the path integral over the quark field:
\begin{align}
\Omega =& -2 N_f\int \frac{d^3{\rm p}}{(2\pi)^3}
         \Bigl[ 3 E ({\rm p}) \notag\\
        & \hspace{-4mm}+ \frac{1}{\beta}
           \ln~ [1 + 3(\Phi+\Phi_{\rm H} e^{-\beta E^-({\bf p})}) 
           e^{-\beta E^-({\bf p})}+ e^{-3\beta E^-({\bf p})}]
         \notag\\
        & \hspace{-4mm}+ \frac{1}{\beta} 
           \ln~ [1 + 3(\Phi_{\rm H}+{\Phi e^{-\beta E^+({\bf p})}}) 
              e^{-\beta E^+({\bf p})}+ e^{-3\beta E^+({\bf p})}]
	      \Bigl]\notag\\
        & \hspace{-4mm}+U+{\cal U} , 
\label{eq:E12} 
\end{align}
with $E({\rm p})=\sqrt{{\bf p}^2+M^2}$, $M=m_0 + \Sigma_{\rm s}$ 
and $E^\pm({\rm p})=E({\rm p})\pm \tilde{\mu}$. 
Here the effective chemical potential $\tilde{\mu}$ is defined by 
$\tilde{\mu} =\mu_{\rm R}+iT\theta - 2G_{\rm v}n$.

We use ${\cal U}$ of Ref.~\cite{Ratti1}: 
\begin{align}
&{{\cal U}\over{T^4}} =  -\frac{b_2(T)}{2} \Phi_{\rm H}\Phi
              -\frac{b_3}{6}({\Phi_{\rm H}}^3+ \Phi^3)
              +\frac{b_4}{4}(\Phi_{\rm H}\Phi)^2, \label{eq:E13} \\
&b_2(T)   = a_0 + a_1\Bigl(\frac{T_0}{T}\Bigr)
                 + a_2\Bigl(\frac{T_0}{T}\Bigr)^2
                 + a_3\Bigl(\frac{T_0}{T}\Bigr)^3.   
\label{eq:E14}
\end{align}
Parameters of ${\cal U}$ are fitted to 
results of lattice simulations in the pure gauge system with 
finite $T$~\cite{Boyd,Kaczmarek}. In the pure gauge system 
the parameter $T_0$ agrees with the critical temperature $T_{\rm D}$ 
of deconfinement phase transition. Hence, as a reasonable choice, 
one can adjust $T_0$ 
to $T_{\rm D}=270$~MeV of pure gauge lattice simulations. 
However, the PNJL model with this value of $T_0$ yields somewhat larger 
value of $T_{\rm D}$ than $\sim173$~MeV predicted 
by the full LQCD simulation~\cite{Karsch3,Karsch4,MCheng}. 
In this paper, we then take the rescaled value 
$T_0=190$~MeV that gives $T_{\rm D}=$176~MeV 
in the PNJL calculation with $\mu=0$~\cite{Sakai}.

As expected, $\Omega$ is 
invariant under the extended ${\mathbb Z}_3$ transformation, 
\begin{align}
&e^{\pm i \theta} \to e^{\pm i \theta} e^{\pm i{2\pi k\over{3}}},\quad  
\Phi(\theta)  \to \Phi(\theta) e^{-i{2\pi k\over{3}}}, 
\notag\\
&\Phi_{\rm H}(\theta) \to \Phi_{\rm H}(\theta) e^{i{2\pi k\over{3}}} .
\label{eq:K2}
\end{align}
Thus, the extended ${\mathbb Z}_3$ invariance of $\Omega$ is held, even if 
the vector-type four-quark interaction is added to the PNJL Langrangian. 
In general, the symmetry persists, even if 
any sort of multi-quark interactions are added to the NJL sector.

As mentioned above, it is essential to introduce the modified Polyakov
loop,
\begin{eqnarray} 
\Psi \equiv e^{i\theta}\Phi,~~~~~\Psi_{\rm H} \equiv
e^{-i\theta}\Phi_{\rm H},
\label{eq:NE11}
\end{eqnarray}
invariant under the the extended ${\mathbb Z}_3$ transformation (\ref{eq:K2}). 
\begin{widetext} 
The extended ${\mathbb Z}_3$ transformation is then 
rewritten into 
\begin{align}
e^{\pm i \theta} \to e^{\pm i \theta} e^{\pm i{2\pi k\over{3}}}, \quad
\Psi(\theta) \to \Psi(\theta), \quad 
\Psi_{\rm H}(\theta) \to \Psi_{\rm H}(\theta) ,
\label{eq:K2d}
\end{align}
and $\Omega$ is also into  
\begin{align}
\Omega =  -2 N_f \int \frac{d^3{\rm p}}{(2\pi)^3}
          \Bigl[ 3 E ({\rm p}) 
          &+ \frac{1}{\beta}\ln~ [1 + 
          3\Psi e^{-\beta E({\bf p})}e^{\beta \hat{\mu}_{n}}
          + 3\Psi_{\rm H}e^{-2\beta E({\bf p})}e^{2 \beta \hat{\mu}_{n}}
          e^{3 i \theta}
          + e^{-3\beta E({\bf p})}e^{3 \beta \hat{\mu}_{n}}
          e^{3 i \theta}]
\notag\\
          &\hspace{-5mm}+ \frac{1}{\beta} 
           \ln~ [1 + 
           3\Psi_{\rm H} e^{-\beta E({\bf p})}e^{- \beta \hat{\mu}_{n}}
          + 3\Psi e^{-2\beta E({\bf p})}e^{-2 \beta \hat{\mu}_{n}}
          e^{-3 i \theta}
          + e^{-3\beta E({\bf p})}e^{-3 \beta \hat{\mu}_{n}}
          e^{-3 i \theta}]
	      \Bigl]+U
\notag\\
          &-{b_2(T)T^4\over{2}}\Psi_{\rm H} \Psi
          -{b_3T^4\over{6}}({\Psi_{\rm H}}^3 e^{3 i \theta}
          +\Psi^3 e^{-3 i \theta})
          +{b_4T^4\over{4}}(\Psi_{\rm H} \Psi)^2,
\label{eq:K3} 
\end{align}
\end{widetext}
where $\hat{\mu}_{n}=\mu_{\rm R}-2 G_{\rm v}n$, 
and the factor $\exp{(\pm i 3 \theta)}$ is also invariant 
under the transformation (\ref{eq:K2d}). 
Variables $X =\Psi, \Psi_{\rm H}, \sigma$ and $n$ are 
determined from the stationary conditions 
\bea
\partial \Omega/\partial X=0 ,
\label{condition}
\eea
by solving the equations for $X$. 
The thermodynamic potential $\Omega(\theta)$ at each $\theta$ 
is obtained by inserting the solutions into (\ref{eq:K3}). 
Such a calculation is possible for any complex $\mu$.

\subsection{RW periodicity and even/odd property}
\label{RW periodcity}

We begin with the case of imaginary chemical potential $\mu=iT\theta$, 
that is, take $\mu_{\rm R}=0$ in (\ref{eq:K3}).
The equations (\ref{condition}) depend on 
$\theta$ only through the factor $\exp{(3 i \theta)}$, 
so that the solutions $X$ are functions of the factor: 
\bequ
X=X(e^{3 i \theta}). 
\label{fx}
\eequ
Since $\Omega(\theta)$ is obtained by inserting the solutions $X$ 
into (\ref{eq:E12}), $\Omega(\theta)$ is also a function 
of $\exp{( 3 i \theta )}$. 
Hence, we get 
\bequ
\Omega(\theta +{\tfrac{2\pi k}{3}})=\Omega(\theta ),~~~
X(\theta +{\tfrac{2\pi k}{3}})=X(\theta ). 
\eequ
Thus, quantities invariant with respect to the extended ${\mathbb Z}_3$ 
transformation have the RW periodicity. It is found 
from the proof shown above that the property is not influenced by 
the presence/absence of the four-quark vector-type interaction. 
This will be confirmed later through Fig.\ref{fig1}. 

Next we discuss their more detailed symmetry properties. 
Taking the complex conjugate to (\ref{eq:K3}), one can find that 
the complex conjugate $\Omega^*$ has the same form as the original one 
$\Omega$, 
if $\sigma^*$, $n^*$, $\Psi^*$ and $\Psi_{\rm H}^*$ are replaced by 
$\sigma$ and $-n$, $\Psi_{\rm H}$ and $\Psi$, respectively. 
This indicates that 
the solutions $X^*$ of $\partial \Omega^*/\partial X^*=0$ are related to 
those $X$ of $\partial \Omega/\partial X=0$ as 
\bequ
\sigma^* = \sigma,~~
n^*=-n, ~~
\Psi^*=\Psi_{\rm H}, ~~
\Psi_{\rm H}^*=\Psi . 
\eequ
The last two equations represent that $\Psi_{\rm H}$ is the complex conjugate 
to $\Psi$, so we simply use $\Psi^*$ instead of $\Psi_{\rm H}$ 
in the case of imaginary $\mu$. Meanwhile, the first two equations indicate 
\bequ
\frac{d\Omega^*}{dm_0}=\frac{d\Omega}{dm_0}, \quad 
\frac{d\Omega^*}{d\theta}=\frac{d\Omega}{d\theta}, 
\label{Omega-relation} 
\eequ
because $\sigma=d\Omega/dm_0$ and 
$n=-d\Omega/d\mu=i\beta d\Omega/d\theta$. 
The relations (\ref{Omega-relation}) are satisfied, only when 
$\Omega$ is real. Thus, $\Omega$ is real, 
so that $n$ is pure imaginary and $\sigma$ is real.

As shown in (\ref{fx}), $n$ 
is a function of $\cos(3\theta)$ and $i \sin(3\theta)$. 
This shows, together with the fact that 
$n$ is pure imaginary, that $n$ is $\theta$-odd 
(odd under the transformation $\theta \to -\theta$), since
\bea
&&-n(\cos(3\theta),i \sin(3\theta)) 
=n(\cos(3\theta),i \sin(3\theta))^*
\nonumber\\
&&=n(\cos(3\theta),-i \sin(3\theta))
=n(\cos(-3\theta),i \sin(-3\theta)) .
\nonumber\\
\eea
Similarly, $\sigma$ is real and 
a function of $\cos(3\theta)$ and $i \sin(3\theta)$. This leads 
to the fact that  $\sigma$ is $\theta$-even, because 
\bea
&&\sigma(\cos(3\theta),i \sin(3\theta)) 
=\sigma(\cos(3\theta),i \sin(3\theta))^*
\nonumber\\
&&=\sigma(\cos(3\theta),-i \sin(3\theta))
=\sigma(\cos(-3\theta),i \sin(-3\theta)).
\nonumber\\
\eea

The potential $\Omega(\theta)$ depends on $\theta$ through 
$\Psi(\theta)$, $\Psi(\theta)^{*}$, $\sigma(\theta)$, $n(\theta)$
and $e^{3i\theta}$. We then denote $\Omega(\theta)$ by  
$\Omega(\theta)=
\Omega(\Psi(\theta),\Psi(\theta)^{*},n(\theta),e^{3i\theta})$, 
where $\sigma(\theta)$ is suppressed since it is $\theta$-even and does not 
make any influence on discussions shown below. 
Equation (\ref{eq:K3}) keeps the same form 
under the transformation $\theta \to -\theta$, 
if $\Psi(\theta)$ and $\Psi(\theta)^{*}$ are replaced by 
$\Psi(-\theta)^{*}$ and $\Psi(-\theta)$, respectively. 
This indicates that 
\bea
\Psi(-\theta)=\Psi(\theta)^{*} \quad {\rm and} \quad 
\Psi(-\theta)^{*}=\Psi(\theta) .
\label{psi-z2}
\eea
Using these properties and the fact that $\Omega$ is real, 
one can show that 
\begin{align}
\Omega(\theta)=&(\Omega(\theta))^*
=\Omega(\Psi(\theta)^{*},\Psi(\theta),-n(\theta),e^{-3i\theta} )
\notag\\
=&\Omega(\Psi(-\theta),\Psi(-\theta)^{*},n(-\theta) ,e^{-3i\theta})
=\Omega(-\theta) .
\end{align}
Therefore, we can summarize that 
$\Omega$ is a periodic even function of $\theta$ with 
a period $2\pi/3$, $\Omega(\theta)=
\Omega(\theta+2 \pi k/3)=\Omega(-\theta)$.  
The chiral condensate $\sigma(\theta)$ is also a 
periodic even function of $\theta$, 
while the quark number density $n$ is 
a periodic odd function of $\theta$. 
These properties are not changed by the presence/absence of 
the vector-type four-quark interaction, as understood from 
the proof shown above.

The real (imaginary) part of $\Psi$ is $\theta$-even ($\theta$-odd), because  
\bea
{\rm Re}[\Psi(\theta)]&=&(\Psi(\theta) + \Psi(\theta)^*)/2=
{\rm Re}[\Psi(-\theta)] ,\nonumber\\
{\rm Im}[\Psi(\theta)]&=&(\Psi(\theta) - \Psi(\theta)^*)/(2i)
=-{\rm Im}[\Psi(-\theta)],
\label{eq:NE23}
\eea
where use has been made of (\ref{psi-z2}). 
Thus, the real (imaginary) part of $\Psi$ is a periodic 
even (odd) function of $\theta$.
Similarly, the absolute value $|\Psi |$ (phase $\phi$) of the Polyakov loop is a periodic even (odd) function of $\theta$, because 
$|\Psi|=\sqrt{({\rm Re }[\Psi] )^2+({\rm Im}[\Psi] )^2}$ 
($\phi =\arctan{({{\rm Im}[\Psi]/{{\rm Re}[\Psi]}})}$). 
Obviously, these properties persist also in the case of $G_{\rm v}=0$.

Since all quantities of our interest have the RW periodicity, 
here we consider a period $0 \le \theta \le 2\pi/3$ in the $\theta$-$T$ plane. 
In the region, periodic even functions such as 
$\Omega(\theta)$, $\sigma(\theta)$, ${\rm Re}[\Psi(\theta)]$ and $|\Psi |$ are 
symmetric with respect to a line $\theta=\pi/3$. 
This indicates that such an even function $X_\mathrm{e}$ has 
a cusp at $\theta=\pi/3$, 
if the gradient $dX_\mathrm{e}/d\theta |_{\theta=\pi/3 \pm \epsilon}$ is not zero, 
where $\epsilon$ is a positive infinitesimal. 
The phase transition appearing  
in $X_\mathrm{e}$ at $\theta=\pi/3$ is of the second order. 
Meanwhile, ${\rm Im}[\Psi(\theta)]$, $\phi$ and $n$ 
are periodic odd functions of $\theta$. 
This leads to the fact that these are discontinuous at $\theta=\pi/3$, 
if the odd functions are not zero there. 
Thus, the phase transitions 
appearing in the odd functions at $\theta=\pi/3$ are of 
the first order. 
These are seen in the high $T$ region, as shown later with numerical calculations. 
The phase transition appearing at $\theta=\pi/3$ is called the 
RW transition. Thus, the RW phase transition is a family of 
first order transitions in $\theta$-odd quantities and 
second order ones in $\theta$-even quantities. 
This result is not changed by the presence/absence of 
vector-type four quark interaction.

As mentioned above, 
$\Omega$ of (\ref{eq:K3}) is invariant 
under the extended ${\mathbb Z}_{3}$ transformation 
for any complex chemical potential $\mu=\mu_{\rm R}+i T \theta$. 
This implies that $X$ has the RW periodicity, 
$X(\mu_{\rm R}+iT \theta)=X(\mu_{\rm R}+i T (\theta+2\pi k/3))$.  
Actually, $\Omega$ depends on $\theta$ only through the factor 
$\exp(3i\theta)$, even if $\mu_{\rm R}$ is nonzero. 
Hence, the solutions $X$ of 
$\partial \Omega/\partial X=0$ are functions 
of $\exp(3i\theta)$. 
Thus, the concept of 
the extended ${\mathbb Z}_{3}$ transformation is useful for 
any complex chemical potential.

Finally, we briefly consider the case of real $\mu$ 
by taking $\theta=0$ in  (\ref{eq:K3}); 
here note that 
$\Psi = \Phi$ and $\Psi_{\rm H} = \Phi_{\rm H}$. 
The complex conjugate potential $\Omega^*$ 
has the same form as $\Omega$, 
if $\sigma^*$, $n^*$, $\Psi^*$ and $\Psi_{\rm H}^*$ are replaced by 
$\sigma$, $n$, $\Psi$ and $\Psi_{\rm H}$, respectively. 
This indicates that 
the solutions $X^*$ of $\partial \Omega^*/\partial X^*=0$ 
are related to the solutions $X$ of $\partial \Omega/\partial X=0$ as 
\bequ
\sigma^* = \sigma,~~
n^*=n, ~~
\Psi^*=\Psi, ~~
\Psi_{\rm H}^*=\Psi_{\rm H} . 
\label{relation-X-2}
\eequ
Thus, the four quantities are real. Furthermore, 
the first and second equations of (\ref{relation-X-2}) 
are reduced to 
\bequ
\frac{d\Omega^*}{dm_0}=\frac{d\Omega}{dm_0}, \quad 
\frac{d\Omega^*}{d\mu_{\rm R}}=\frac{d\Omega}{d\mu_{\rm R}},  
\label{Omega-relation-2} 
\eequ
respectively. This indicates that $\Omega$ is also real.

Equation (\ref{eq:K3}) keeps the same form 
under the transformation $\mu_{\rm R} \to -\mu_{\rm R}$, 
if $\Psi$, $\Psi_{\rm H}$, $\sigma$ 
and $n$ 
are replaced by 
$\Psi_{\rm H}$, $\Psi$, 
$\sigma$ and $-n$, 
respectively. 
This indicates that 
the solutions $X(-\mu_{\rm R})$ of 
$\partial \Omega(-\mu_{\rm R})/\partial X=0$ 
are related to the solutions $X(\mu_{\rm R})$ of 
$\partial \Omega(\mu_{\rm R})/\partial X=0$ as 
\bea
\Psi_{\rm H}(-\mu_{\rm R})&=&\Psi(\mu_{\rm R}), \quad
\Psi(-\mu_{\rm R})=\Psi_{\rm H}(\mu_{\rm R}),
\nonumber \\
\sigma(-\mu_{\rm R})&=&\sigma(\mu_{\rm R}), \quad
n(-\mu_{\rm R})=-n(\mu_{\rm R}) .
\label{psi-z2-2}
\eea
Using (\ref{psi-z2-2}) in (\ref{eq:K3}), we can easily confirm that 
$\Omega(\mu_{\rm R})=\Omega(-\mu_{\rm R})$. 
Thus, $\Omega$ and $\sigma$ are even functions of $\mu$ 
in both the real and imaginary $\mu$ regions, so 
the chiral transition curve shown later is plotted in the $\mu^2$-$T$ plane.

\section{Numerical results}
\label{Numerical-results}

We proceed to numerical calculations. 
Since the PNJL model are nonrenormalizable, it is then needed to 
introduce a cutoff in the momentum integration. 
Here we take a three-dimensional momentum cutoff $\Lambda$. 
Hence, the present model has four parameters 
$m_0$, $\Lambda$, $G_{\rm s}$, $G_{\rm v}$ in the NJL sector. 
Following Ref.~\cite{Kashiwa}, we take $m_0=5.5$~MeV, 
$\Lambda =0.6315$ GeV and $G_{\rm s}=5.498$~GeV$^{-2}$ 
that reproduce the pion decay constant $f_\pi =93.3$~MeV 
and the pion mass $M_\pi =138$~MeV. 
That is, we pin down parameters other than $G_{\rm v}$ at $\mu=0$, then 
we treat $G_{\rm v}$ as a free parameter and vary its value in the range 
$0 \le G_{\rm v} \le G_{\rm s}$.

\begin{figure}[htbp]
\begin{center}
 \includegraphics[width=0.3\textwidth,angle=-90]{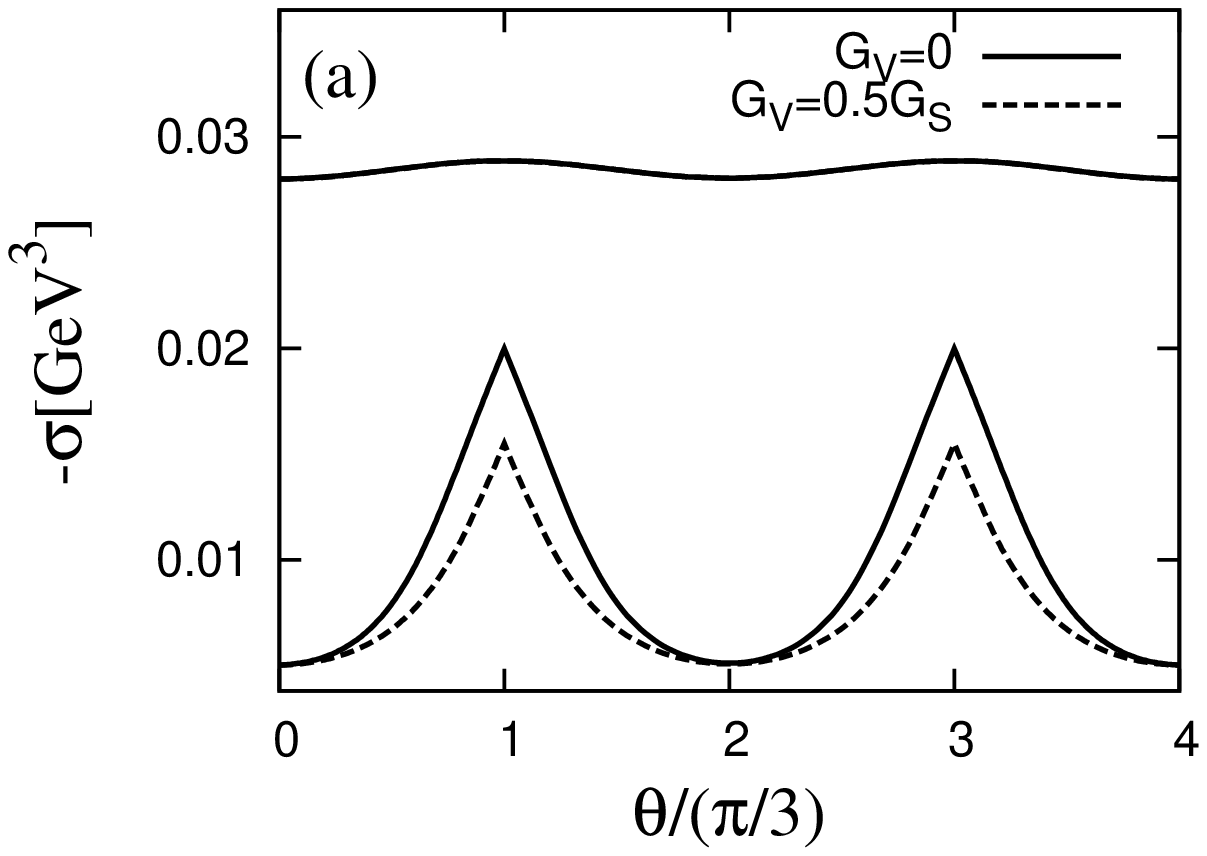} 
 \includegraphics[width=0.3\textwidth,angle=-90]{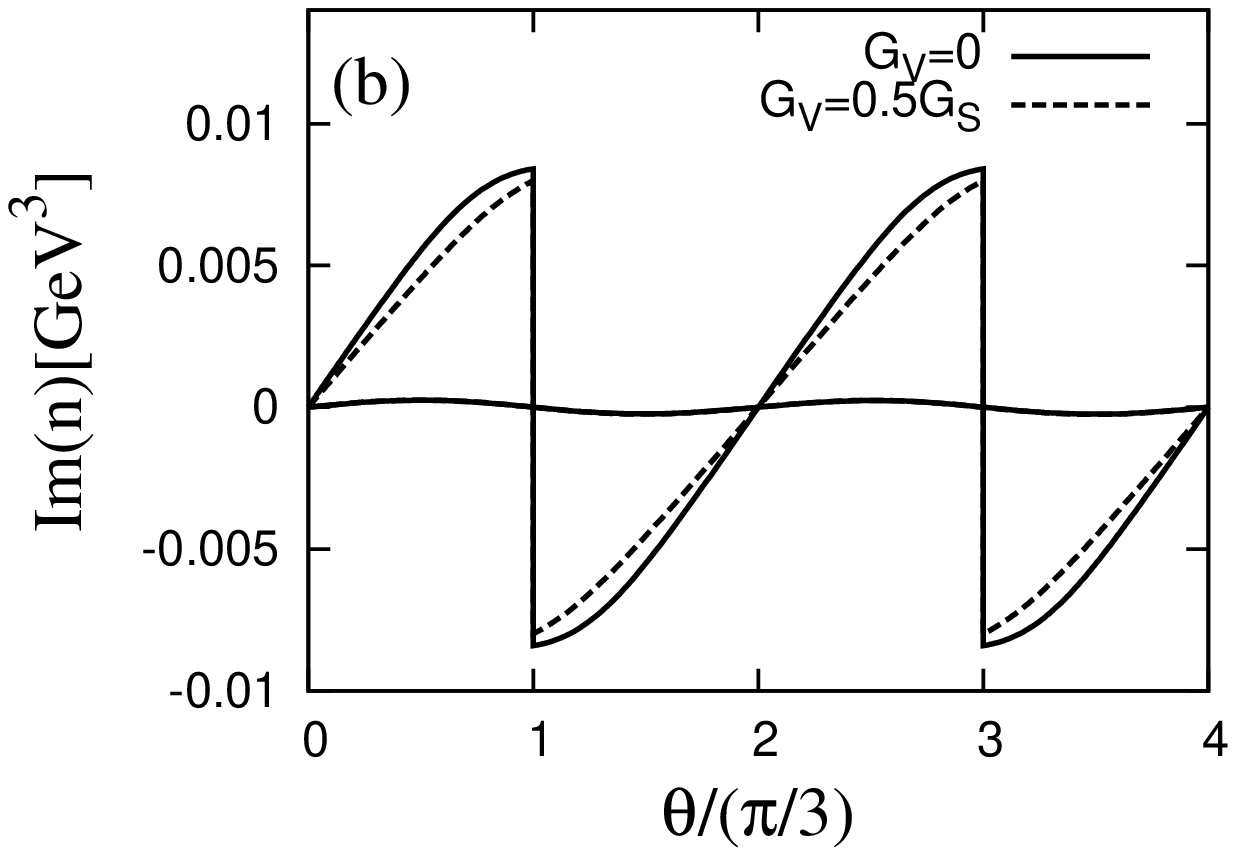} 
 \includegraphics[width=0.3\textwidth,angle=-90]{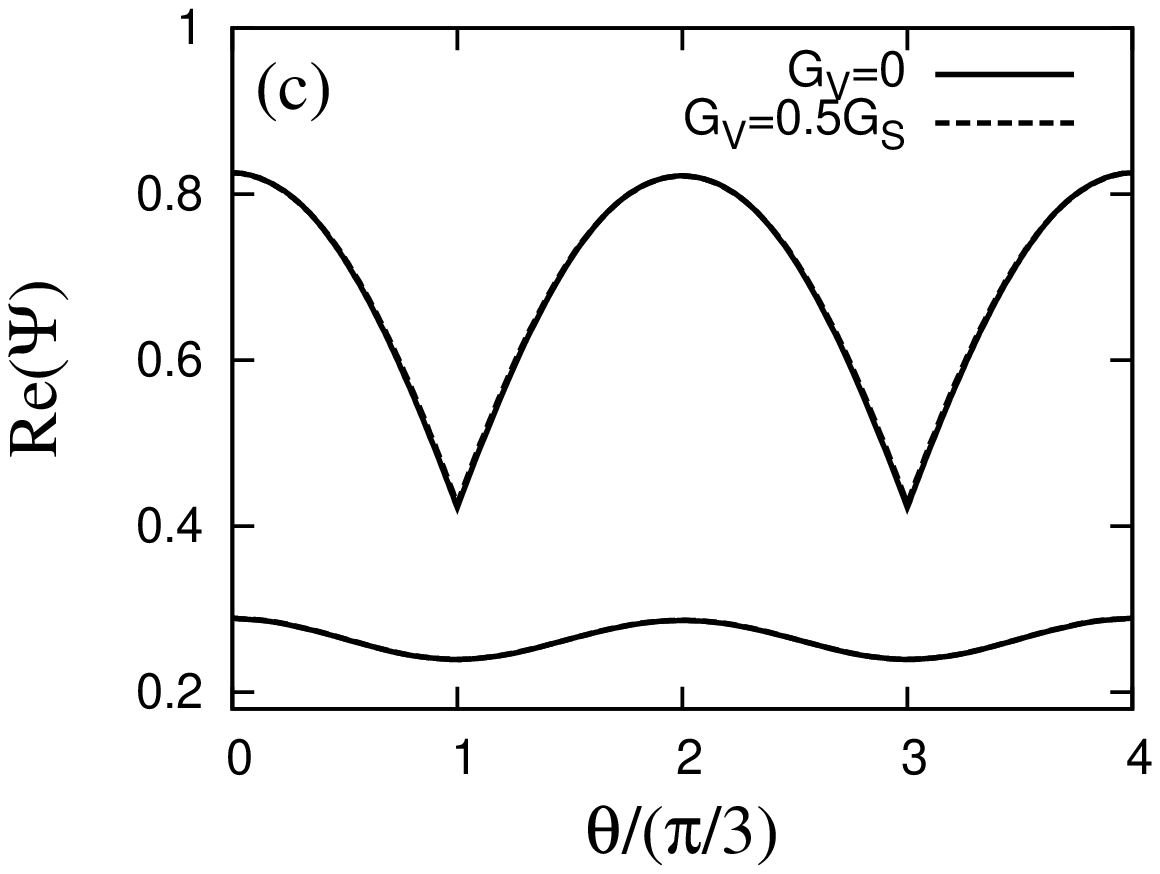} 
 \includegraphics[width=0.3\textwidth,angle=-90]{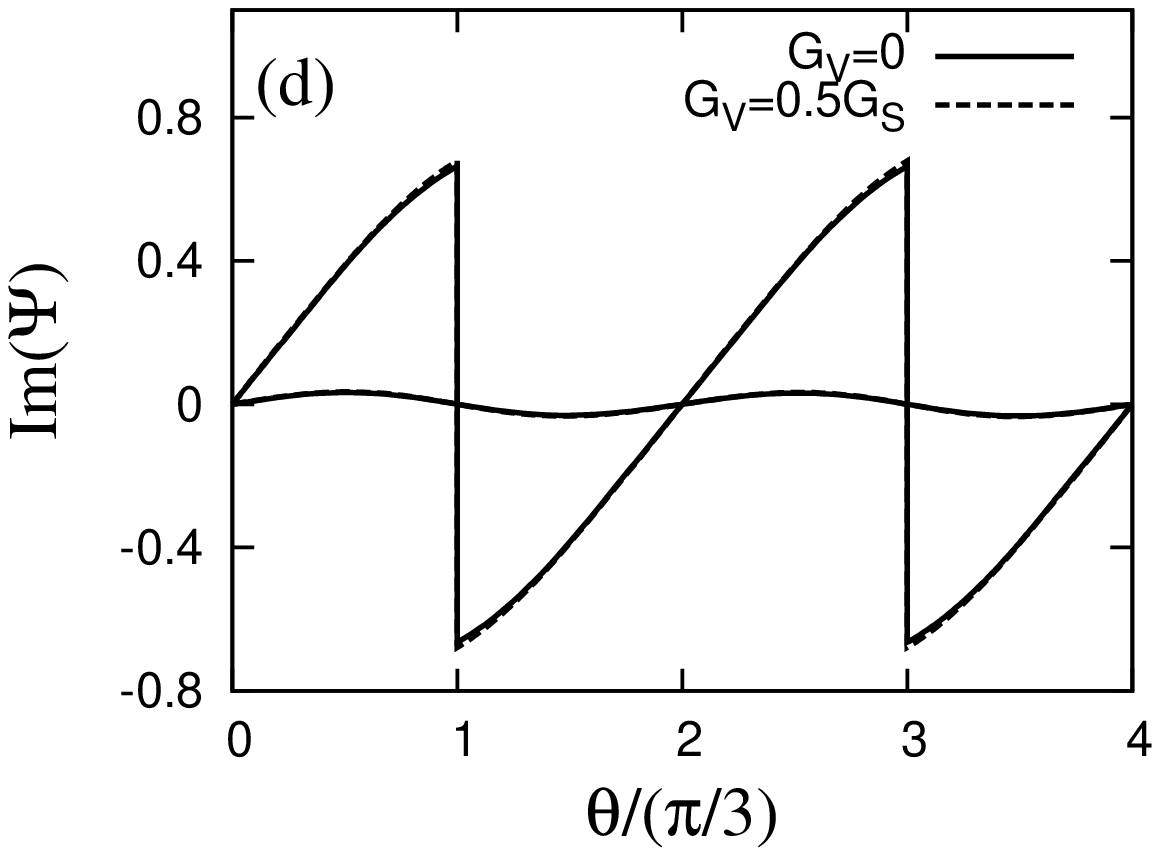} 
\end{center}
\caption{The Roberge-Weiss periodicity of (a) chiral condensate, 
(b) imaginary part of the quark number density, (c) real part 
of the modified Polyakov loop, and (d) its imaginary part. 
The solid curves are for $G_{\rm v}=0$ and the dashed ones are for 
$G_{\rm v}=0.5G_{\rm s}$. The curves with cusps ($\theta$-even 
quantities) and those with discontinuities ($\theta$-odd quantities) 
are for $T=250$~MeV, while smooth ones are for $T=170$~MeV. }
\label{fig1}
\end{figure}

Figure~\ref{fig1} represents the RW periodicity of various quantities 
calculated 
with/without the vector-type interaction. These graphs clearly shows that the 
RW periodicity is not affected by the inclusion of the vector-type interaction 
as argued analytically. 
At $T=170$~MeV, 
which is below 
the endpoint of the RW phase transition (point A of Fig.~\ref{fig3}), 
all quantities are smooth as functions of $\theta$, whereas at $T=250$~MeV, 
which is above the endpoint, 
the $\theta$-even quantities, $\sigma$ and ${\rm Re}[\Psi]$, 
exhibit second-order and the $\theta$-odd quantities, ${\rm Im}[n]$ and 
${\rm Im}[\Psi]$, do first-order phase transitions. 
We note here that the present endpoint A is lower than 
in our previous 
works~\cite{Sakai} because $T_0$ in the Polyakov potential was rescaled in the 
present calculation.

\begin{figure}[htbp]
\begin{center}
 \includegraphics[width=0.3\textwidth,angle=-90]{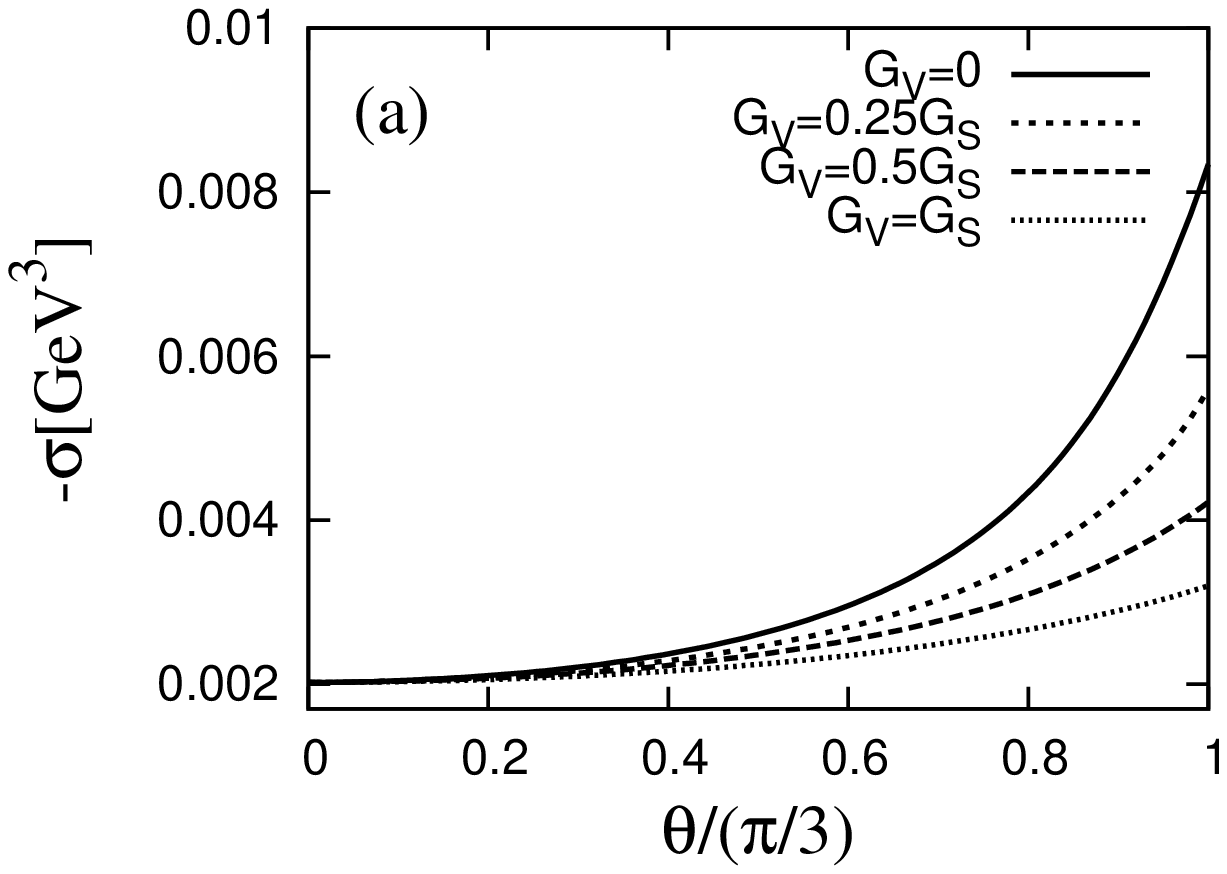} 
 \includegraphics[width=0.3\textwidth,angle=-90]{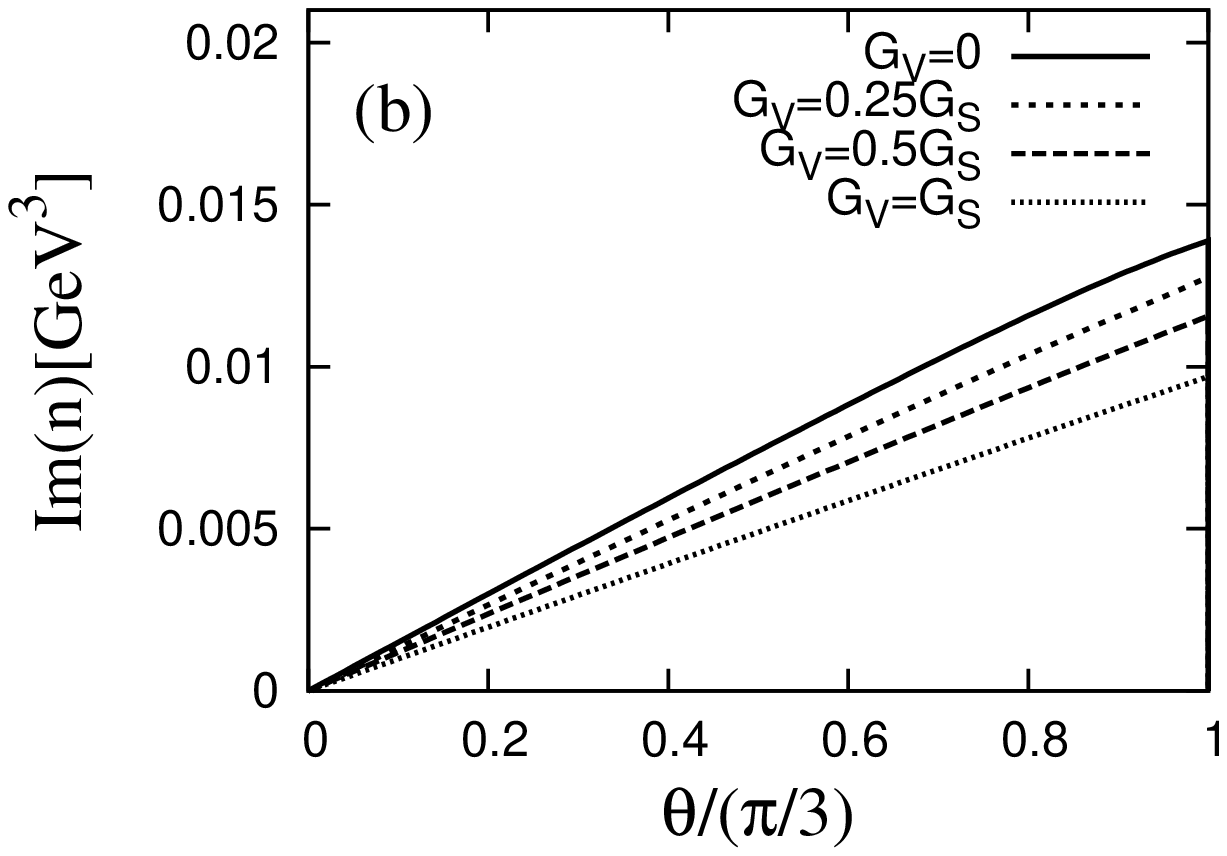} 
 \includegraphics[width=0.3\textwidth,angle=-90]{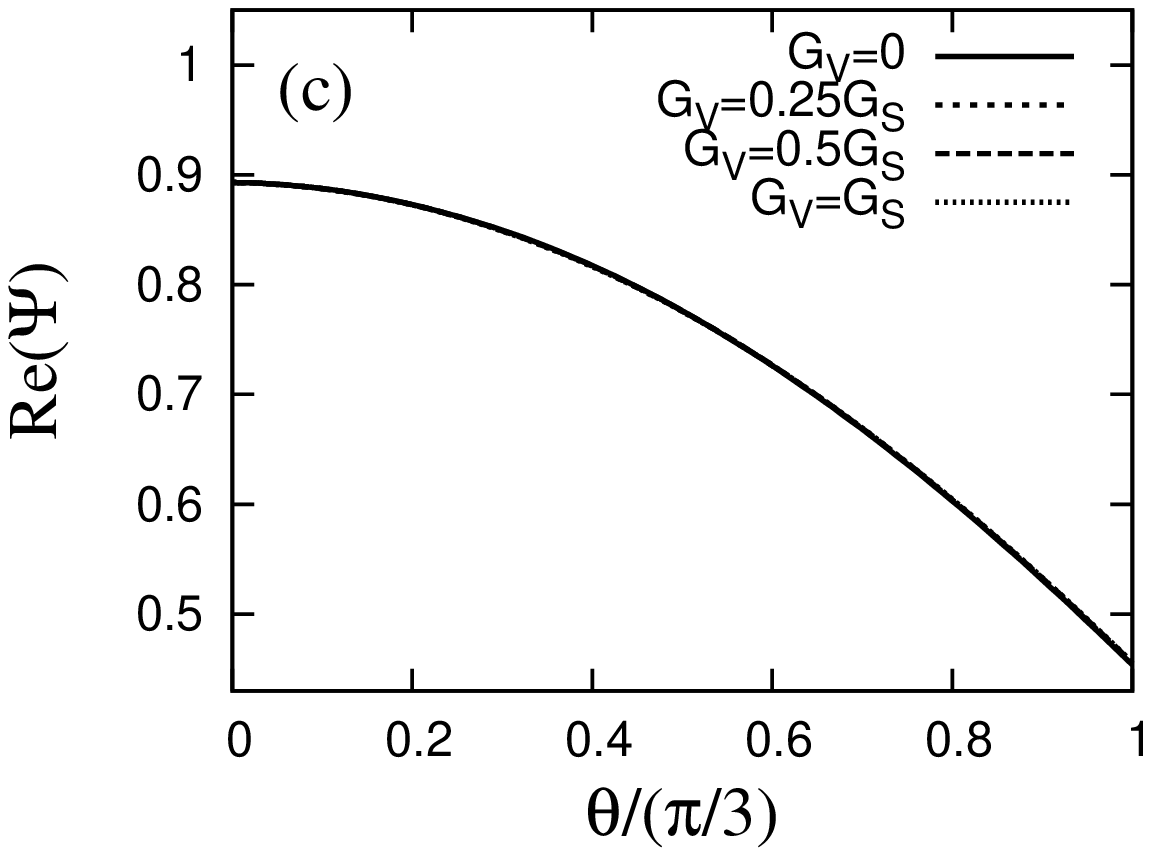} 
 \includegraphics[width=0.3\textwidth,angle=-90]{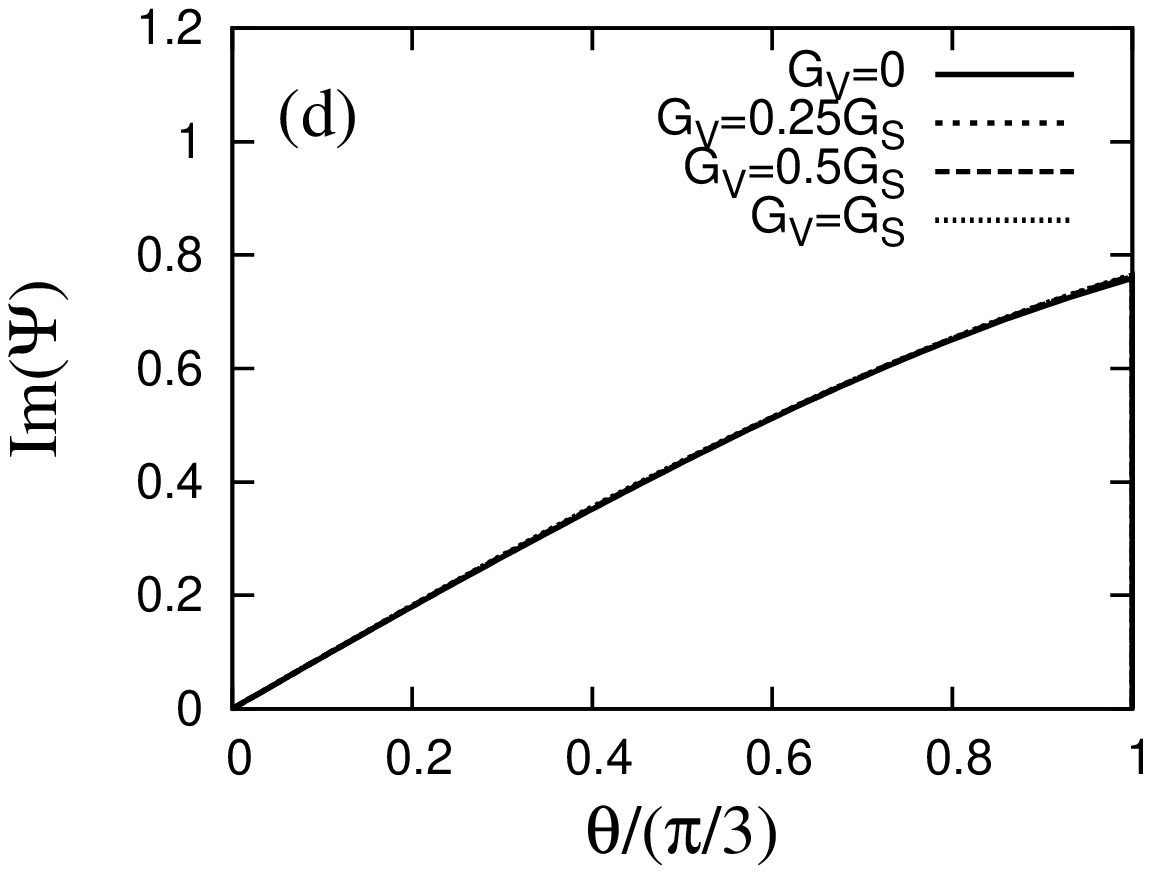} 
\end{center}
\caption{The impact of the vector-type four-quark interaction on the 
four quantities same as in Fig.~\ref{fig1} at $T=300$~MeV for a half 
Roberge-Weiss period. }
\label{fig2}
\end{figure}

\begin{figure}[htbp]
\begin{center}
 \includegraphics[width=0.3\textwidth,angle=-90]{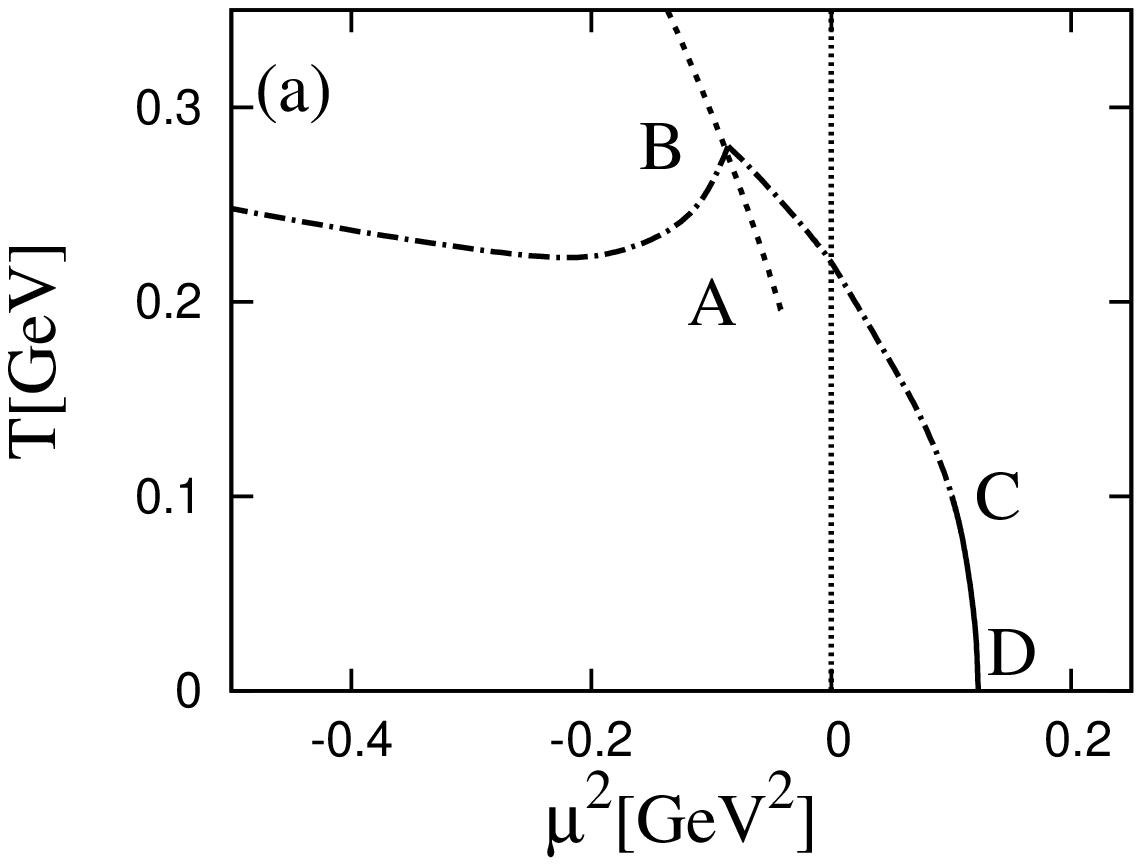}
 \includegraphics[width=0.3\textwidth,angle=-90]{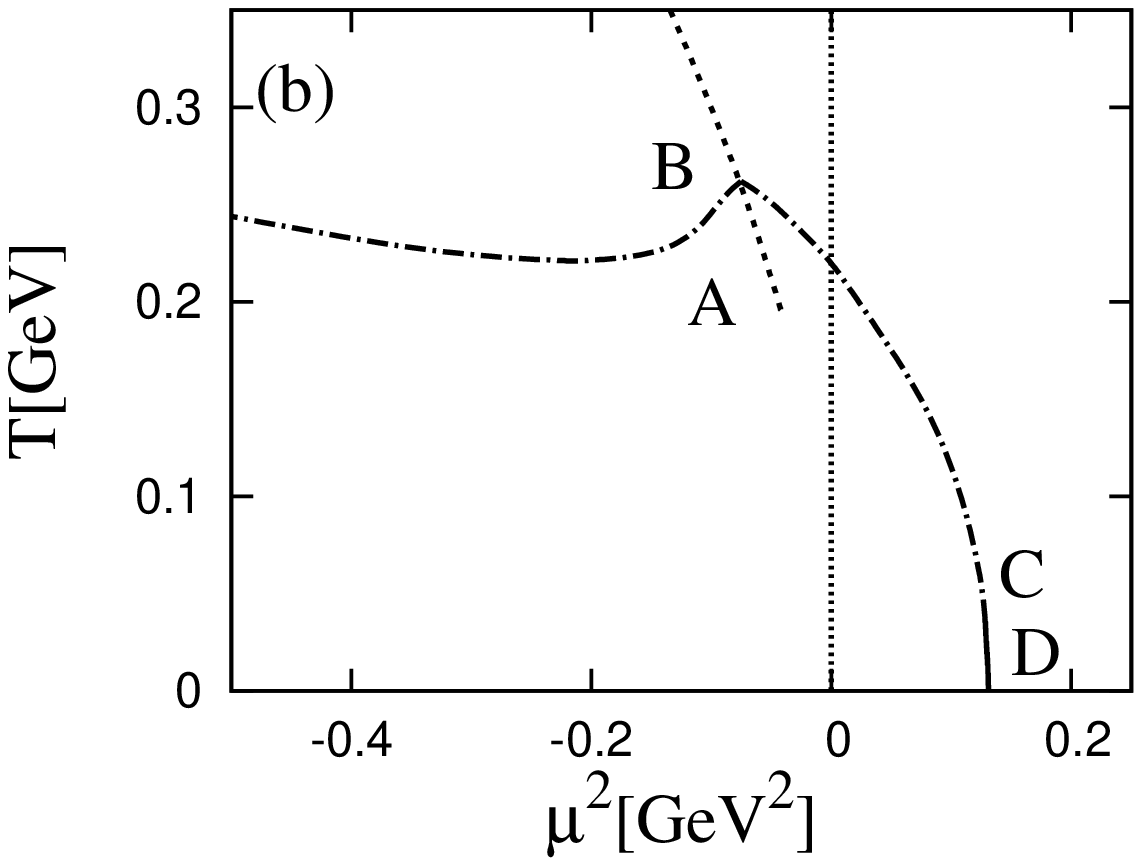}
\end{center}
\caption{
The phase diagram in the $\mu^2$-$T$ plane, (a) for 
$G_{\rm v}=0$ and (b) for $G_{\rm v}=0.25 G_{\rm s}$. 
The point A is the endpoint of the RW phase transition, 
B is the crossing point of the ordinary chiral crossover transition 
curve and the RW phase transition line, 
C is the chiral critical endpoint, and 
D is the chiral transition point at $T=0$.}
\label{fig3}
\end{figure}

Next we pay attention to the impact of the vector-type interaction on various 
quantities. 
Figure \ref{fig2} represents the four quantities same as in Fig.~\ref{fig1} 
for half a RW period, $0 \le \theta \le \pi/3$, calculated with 
various strengths $G_\mathrm{v}$. 
Note that here we present the results at $T=300$ MeV since the effect of the 
vector-type interaction is more conspicuous at higher temperatures. 
First of all, the chiral condensate shown in Fig.~\ref{fig2}(a), in particular 
around the RW transition at $\theta=\pi/3$, 
is sensitive to $G_\mathrm{v}$. The calculated result means that the 
vector-type interaction reduces the constituent quark mass $M$. 
The vector-type interaction suppresses the 
effective chemical potential 
$\tilde{\mu}=i\theta T - 2i G_{\rm v}{\rm Im}[n]$. 
This directly affects ${\rm Im}[n]$, namely, ${\rm Im}[n]$ decreases as 
$G_\mathrm{v}$ increases. Although the indirect effect through $M$, mentioned 
just above, works oppositely, the direct effect is stronger and therefore 
survives. Consequently effects of the vector-type interaction on 
${\rm Im}[n]$ shown in Fig.~\ref{fig2}(b) as well as on $\sigma$ 
are significant. 
Figures \ref{fig2}(c) and (d) indicate that the effects 
on the modified Polyakov 
loop are negligible because it is fully determined by the Polyakov potential. 
The physical value of $G_\mathrm{v}$ is still an open matter although its 
effect is significant in the physical real-$\mu$ phase diagram 
(see Fig.~\ref{fig4} below). Since the present calculation shows that its 
impacts on the chiral condensate and the quark number density are appreciable 
in the imaginary-$\mu$ 
region where both lattice simulations and effective model calculations are 
available, we expect that it would be possible to determine a physical 
strength of $G_{\rm v}$ 
by comparing both the results, although 
the former is not available yet for the case of the two flavor QCD. 

We report the impact of the vector-type four-quark interaction on the 
$\mu^2$-$T$ phase diagram ranging from negative to positive $\mu^2$ in 
Fig.~\ref{fig3}, since the order parameters 
$\sigma$ are functions of $\mu^2$ 
as proved analytically above. 
This figure indicates that effect of the vector-type interaction on the 
phase diagram is visible in the real-$\mu$ part; it has been 
known that the critical endpoint (CEP; point C) moves to lower $T$ and higher 
$\mu$ as $G_{\rm v}$ increases and eventually disappears at 
$G_{\rm v} \approx 0.38 G_{\rm s}$~\cite{KKKN,Kashiwa,Kashiwa3,Fukushima2}. 
In contrast, its effects are less in the imaginary-$\mu$ part 
of the phase diagram although the values of $\sigma$ and ${\rm Im}[n]$ are 
sensitive to it. 
This can be understood from the fact that the effect of finite $G_{\rm v}$ 
on $\tilde\mu$ is important when $T$ is low whereas the RW transition 
occurs at higher $T$. 
We note here that this figure graphs the chiral transition curve although 
the deconfinement transition temperature in the imaginary-$\mu$ is lower. 
We confirmed that their difference decreases as $\theta$ 
increases due to finite $G_{\rm v}$. 

Finally, in order to look into the correlation between the key 
quantities in the imaginary- and real-$\mu$ regions, we present in 
Fig.~\ref{fig4} the relation between the chiral condensate 
at $\theta=\pi/3-\epsilon$ and $T=250$~MeV in the imaginary-$\mu$ region, 
which can be directly measured in lattice simulations, 
and the position of the chiral CEP in the real-$\mu$ region, 
which cannot be measured. Both of them are functions of $G_{\rm v}$; 
here $\epsilon$ is a positive infinitesimal. 
The larger $G_{\rm v}$ is adopted, the smaller $-\sigma$ becomes. 
The right-most point corresponds to $G_{\rm v}=0$ while the left-most 
does to $G_{\rm v}=0.30 G_{\rm s}$. 
Since the correlation shown in Fig.~\ref{fig4} is evident, 
we hope this serves to determine the physical value of $G_{\rm v}$. 

\begin{figure}[htbp]
\begin{center}
 \includegraphics[width=0.25\textwidth,angle=-90]{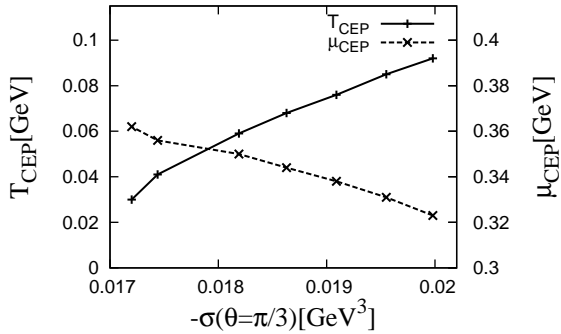}
\end{center}
\caption{
The correlation between the chiral condensate at 
$\theta=\pi/3-\epsilon$ in the 
imaginary-$\mu$ region and the position of the chiral critical endpoint 
(CEP) in the real-$\mu$ region, where $\epsilon$ is a positive infinitesimal. 
These curves correspond to $0\le G_{\rm v}\le 0.30 G_{\rm s}$. 
}
\label{fig4}
\end{figure}

\section{Summary}
\label{Summary}

A key feature of QCD in the imaginary chemical potential 
($\mu$) region is that the partition function has 
the extended ${\mathbb Z}_{3}$ symmetry. 
We prove that the Polyakov-loop extended Nambu--Jona-Lasinio 
(PNJL) model possesses the symmetry, 
even if the vector-type four-quark interaction is newly 
added to the NJL sector of the PNJL Lagrangian. 
Thus, PNJL is a suitable model to analyze not only 
the real $\mu$ region but also the imaginary one. 

Using PNJL, we have analyzed the effect of 
the vector-type four-quark interaction 
in the imaginary $\mu$ region and investigated 
the correlation between the position of the critical endpoint (CEP) in the 
real $\mu$ region and several quantities in the imaginary 
$\mu$ region. 
In the real $\mu$ region, 
the position of CEP is known to be sensitive to the strength of 
the vector-type four-quark interaction. 
Meanwhile, the present analysis shows that in the imaginary $\mu$ region 
the vector-type four-quark interaction largely changes values of 
the chiral condensate and the quark-number density. 
This indicates that these quantities 
in the imaginary $\mu$ region are suitable to 
determine the strength and therefore the QCD phase diagram in the 
real $\mu$ region.

\noindent
\begin{acknowledgments}
The authors thank A. Nakamura for useful discussions and suggestions. 
K.K. and H.K. also thank M. Imachi, H. Yoneyama and M. Tachibana for useful discussions. 
This work has been supported in part 
by the Grants-in-Aid for Scientific Research 
(18540280) of Education, Science, Sports, and Culture of Japan.
\end{acknowledgments}



\begin{thebibliography}{19}
\expandafter\ifx\csname natexlab\endcsname\relax\def\natexlab#1{#1}\fi
\expandafter\ifx\csname bibnamefont\endcsname\relax
  \def\bibnamefont#1{#1}\fi
\expandafter\ifx\csname bibfnamefont\endcsname\relax
  \def\bibfnamefont#1{#1}\fi
\expandafter\ifx\csname citenamefont\endcsname\relax
  \def\citenamefont#1{#1}\fi
\expandafter\ifx\csname url\endcsname\relax
  \def\url#1{\texttt{#1}}\fi
\expandafter\ifx\csname urlprefix\endcsname\relax\def\urlprefix{URL }\fi
\providecommand{\bibinfo}[2]{#2}
\providecommand{\eprint}[2][]{\url{#2}}

%

\bibitem[{\citenamefont{Kogut et~al.}(1983)\citenamefont{Kogut, Stone, {H. W.
  Wyld}, {W. R. Gibbs}, Shigemitsu, {S. H. Shenker}, and {D. K.
  Sinclair}}}]{Kog}
\bibinfo{author}{\bibfnamefont{J.}~\bibnamefont{Kogut}},
  \bibinfo{author}{\bibfnamefont{M.}~\bibnamefont{Stone}},
  \bibinfo{author}{\bibnamefont{{H. W. Wyld}}},
  \bibinfo{author}{\bibnamefont{{W. R. Gibbs}}},
  \bibinfo{author}{\bibfnamefont{J.}~\bibnamefont{Shigemitsu}},
  \bibinfo{author}{\bibnamefont{{S. H. Shenker}}}, \bibnamefont{and}
  \bibinfo{author}{\bibnamefont{{D. K. Sinclair}}}, \bibinfo{journal}{Phys.\
  Rev.\ Lett.} \textbf{\bibinfo{volume}{50}}, \bibinfo{pages}{393}
  (\bibinfo{year}{1983}).

\bibitem[{\citenamefont{Kogut}(2007)}]{Kogut2}
\bibinfo{author}{\bibfnamefont{J.}~\bibnamefont{B.}~\bibnamefont{Kogut}} 
\bibnamefont{and} 
\bibinfo{author}{\bibfnamefont{D.}~\bibnamefont{K.}~\bibnamefont{Sinclair}}  
\bibinfo{journal}{Phys. Rev.\  D} \textbf{\bibinfo{volume}{77}},
\bibinfo{pages}{114503} (\bibinfo{year}{2008}).

\bibitem[{\citenamefont{Forcrand and Philipsen}(2002)}]{FP}
\bibinfo{author}{\bibfnamefont{P.}~\bibnamefont{de}~\bibnamefont{Forcrand}} 
\bibnamefont{and}
\bibinfo{author}{\bibfnamefont{O.}~\bibnamefont{Philipsen}},  
\bibinfo{journal}{Nucl. Phys. } \textbf{\bibinfo{volume}{B642}},
\bibinfo{pages}{290} (\bibinfo{year}{2002});
\bibinfo{author}{\bibfnamefont{P.}~\bibnamefont{de}~\bibnamefont{Forcrand}} 
\bibnamefont{and}
\bibinfo{author}{\bibfnamefont{O.}~\bibnamefont{Philipsen}},  
\bibinfo{journal}{Nucl. Phys. } \textbf{\bibinfo{volume}{B673}},
\bibinfo{pages}{170} (\bibinfo{year}{2003}). 

\bibitem[{\citenamefont{Elia and Lombardo}(2003)}]{Elia}
\bibinfo{author}{\bibfnamefont{M.}~\bibnamefont{D'Elia}} \bibnamefont{and}
\bibinfo{author}{\bibfnamefont{M.}~\bibfnamefont{P.}~\bibnamefont{Lombardo}},  
\bibinfo{journal}{Phys. Rev.\  D} \textbf{\bibinfo{volume}{67}},
\bibinfo{pages}{014505} (\bibinfo{year}{2003});
\bibinfo{author}{\bibfnamefont{M.}~\bibnamefont{D'Elia}} \bibnamefont{and}
\bibinfo{author}{\bibfnamefont{M.}~\bibfnamefont{P.}~\bibnamefont{Lombardo}},  
\bibinfo{journal}{Phys. Rev.\ D} \textbf{\bibinfo{volume}{70}},
\bibinfo{pages}{074509} (\bibinfo{year}{2004});
\bibinfo{author}{\bibfnamefont{M.}~\bibnamefont{D'Elia}},
 \bibinfo{author}{\bibfnamefont{F.}~\bibfnamefont{D.}~\bibnamefont{Renzo}},
\bibnamefont{and}
\bibinfo{author}{\bibfnamefont{M.}~\bibfnamefont{P.}~\bibnamefont{Lombardo}},  
\bibinfo{journal}{Phys. Rev.\ D} \textbf{\bibinfo{volume}{76}},
\bibinfo{pages}{114509} (\bibinfo{year}{2007});
\bibinfo{author}{\bibfnamefont{M.}~\bibfnamefont{P.}~\bibnamefont{Lombardo}}, 
\bibinfo{howpublished}{arXiv:hep-lat/0612017} (\bibinfo{year}{2006}). 

\bibitem[{\citenamefont{Chen and Luo}(2005)}]{Chen}
\bibinfo{author}{\bibfnamefont{H.}~\bibfnamefont{S.}~\bibnamefont{Chen}} \bibnamefont{and}
\bibinfo{author}{\bibfnamefont{X.}~\bibfnamefont{Q.}~\bibnamefont{Luo}},  
\bibinfo{journal}{Phys. Rev.} \textbf{\bibinfo{volume}{D72}},
\bibinfo{pages}{034504} (\bibinfo{year}{2005});
\bibinfo{author}{\bibfnamefont{L.}~\bibfnamefont{K.}~\bibnamefont{Wu}}, 
\bibinfo{author}{\bibfnamefont{X.}~\bibfnamefont{Q.}~\bibnamefont{Luo}},  
\bibnamefont{and}
\bibinfo{author}{\bibfnamefont{H.}~\bibfnamefont{S.}~\bibnamefont{Chen}}, 
\bibinfo{journal}{Phys. Rev.} \textbf{\bibinfo{volume}{D76}},
\bibinfo{pages}{034505} (\bibinfo{year}{2007}).  


\bibitem[{\citenamefont{Sakai et al}(2008)}]{Sakai}
\bibinfo{author}{\bibfnamefont{Y.}~\bibnamefont{Sakai}},
\bibinfo{author}{\bibfnamefont{K.}~\bibnamefont{Kashiwa}}, 
\bibinfo{author}{\bibfnamefont{H.}~\bibnamefont{Kouno}}, 
\bibnamefont{and}
\bibinfo{author}{\bibfnamefont{M.}~\bibnamefont{Yahiro}},
  \bibinfo{journal}{Phys.\ Rev.\  D} \textbf{\bibinfo{volume}{77}},
  \bibinfo{pages}{051901} (\bibinfo{year}{2008});
  \bibinfo{howpublished}{arXiv:hep-ph/0803.1902} (\bibinfo{year}{2008}), 
  to be published in Phys. Rev. D. 

\bibitem[{\citenamefont{Roberge and Weiss}(1986)}]{RW}
\bibinfo{author}{\bibfnamefont{A.}~\bibnamefont{Roberge}} \bibnamefont{and}
\bibinfo{author}{\bibfnamefont{N.}~\bibnamefont{Weiss}},  
\bibinfo{journal}{Nucl. Phys. } \textbf{\bibinfo{volume}{B275}},
\bibinfo{pages}{734} (\bibinfo{year}{1986}). 

\bibitem[{\citenamefont{Nambu and Jona-Lasinio}(1961{\natexlab{a}})}]{NJ1}
\bibinfo{author}{\bibfnamefont{Y.}~\bibnamefont{Nambu}} \bibnamefont{and}
  \bibinfo{author}{\bibfnamefont{G.}~\bibnamefont{Jona-Lasinio}},
  \bibinfo{journal}{Phys.\ Rev.} \textbf{\bibinfo{volume}{122}},
  \bibinfo{pages}{345} (\bibinfo{year}{1961}); 
  \bibinfo{journal}{Phys.\ Rev.} \textbf{\bibinfo{volume}{124}},
  \bibinfo{pages}{246} (\bibinfo{year}{1961}).

\bibitem[{\citenamefont{Meisinger et al.}(1996)}]{Meisinger}
\bibinfo{author}{\bibfnamefont{P.}~\bibnamefont{N.}}~\bibnamefont{Meisinger},
\bibnamefont{and}
\bibinfo{author}{\bibfnamefont{M.}~\bibnamefont{C.}}~\bibnamefont{Ogilvie},  
  \bibinfo{journal}{Phys. Lett.\ B} \textbf{\bibinfo{volume}{379}},
  \bibinfo{pages}{163} (\bibinfo{year}{1996}). 

\bibitem[{\citenamefont{Dumitru}(2002)}]{Dumitru}
\bibinfo{author}{\bibfnamefont{A.}~\bibnamefont{Dumitru}},
\bibnamefont{and}
\bibinfo{author}{\bibfnamefont{R.}~\bibfnamefont{D.}~\bibnamefont{Pisarski}},  
\bibinfo{journal}{Phys.\ Rev.\  D} \textbf{\bibinfo{volume}{66}},
\bibinfo{pages}{096003} (\bibinfo{year}{2002}); 
\bibinfo{author}{\bibfnamefont{A.}~\bibnamefont{Dumitru}},
\bibinfo{author}{\bibfnamefont{Y.}~\bibnamefont{Hatta}},
\bibinfo{author}{\bibfnamefont{J.}~\bibnamefont{Lenaghan}},
\bibinfo{author}{\bibfnamefont{K.}~\bibnamefont{Orginos}},
\bibnamefont{and}
\bibinfo{author}{\bibfnamefont{R.}~\bibfnamefont{D.}~\bibnamefont{Pisarski}},  
\bibinfo{journal}{Phys.\ Rev.\  D} \textbf{\bibinfo{volume}{70}},
\bibinfo{pages}{034511} (\bibinfo{year}{2004}); 
\bibinfo{author}{\bibfnamefont{A.}~\bibnamefont{Dumitru}},
\bibinfo{author}{\bibfnamefont{R.}~\bibfnamefont{D.}~\bibnamefont{Pisarski}},  
\bibnamefont{and}
\bibinfo{author}{\bibfnamefont{D.}~\bibnamefont{Zschiesche}},  
\bibinfo{journal}{Phys.\ Rev.\  D} \textbf{\bibinfo{volume}{72}},
\bibinfo{pages}{065008} (\bibinfo{year}{2005}).

\bibitem[{\citenamefont{Fukushima}(2004)}]{Fukushima}
\bibinfo{author}{\bibfnamefont{K.}~\bibnamefont{Fukushima}}, 
  \bibinfo{journal}{Phys. Lett.\ B} \textbf{\bibinfo{volume}{591}},
  \bibinfo{pages}{277} (\bibinfo{year}{2004}). 

\bibitem[{\citenamefont{{S. K. Ghosh} et al.}(2006)}]{Ghos}
\bibinfo{author}{\bibnamefont{{S. K. Ghosh}}},
  \bibinfo{author}{\bibnamefont{{T. K. Mukherjee}}},
  \bibinfo{author}{\bibnamefont{{M. G. Mustafa}}}, \bibnamefont{and}
  \bibinfo{author}{\bibfnamefont{R.}~\bibnamefont{Ray}},
  \bibinfo{journal}{Phys.\ Rev.\ D} \textbf{\bibinfo{volume}{73}},
  \bibinfo{pages}{114007} (\bibinfo{year}{2006}). 

\bibitem[{\citenamefont{Megias et al.}(2006)}]{Megias}
\bibinfo{author}{\bibfnamefont{E.}~\bibnamefont{Meg{$\acute{\i}$}as}},
\bibinfo{author}{\bibfnamefont{E.}~\bibnamefont{R.}~\bibnamefont{Arriola}},
\bibnamefont{and}
\bibinfo{author}{\bibfnamefont{L.}~\bibnamefont{L.}~\bibnamefont{Salcedo}},  
  \bibinfo{journal}{Phys. Rev.\ D} \textbf{\bibinfo{volume}{74}},
  \bibinfo{pages}{065005} (\bibinfo{year}{2006}). 

\bibitem[{\citenamefont{Ratti et al.}(2006)}]{Ratti1}
\bibinfo{author}{\bibfnamefont{C.}~\bibnamefont{Ratti}},
\bibinfo{author}{\bibfnamefont{M.}~\bibfnamefont{A.}~\bibnamefont{Thaler}},
\bibnamefont{and}
\bibinfo{author}{\bibfnamefont{W.}~\bibnamefont{Weise}},  
  \bibinfo{journal}{Phys. Rev.\ D} \textbf{\bibinfo{volume}{73}},
  \bibinfo{pages}{014019} (\bibinfo{year}{2006}). 

\bibitem[{\citenamefont{Ciminale}(2007)}]{Ciminale}
\bibinfo{author}{\bibfnamefont{M.}~\bibnamefont{Ciminale}},
\bibinfo{author}{\bibfnamefont{R.}~\bibnamefont{Gatto}},
\bibinfo{author}{\bibfnamefont{N.}~\bibfnamefont{D.}~\bibnamefont{Ippolito}},
\bibinfo{author}{\bibfnamefont{G.}~\bibnamefont{Nardulli}},  
\bibnamefont{and}
\bibinfo{author}{\bibfnamefont{M.}~\bibnamefont{Ruggieri}},  
  \bibinfo{journal}{Phys. Rev.\ D} \textbf{\bibinfo{volume}{77}},
  \bibinfo{pages}{054023} (\bibinfo{year}{2008});
\bibinfo{author}{\bibfnamefont{M.}~\bibnamefont{Ciminale}},
\bibinfo{author}{\bibfnamefont{G.}~\bibnamefont{Nardulli}},  
\bibinfo{author}{\bibfnamefont{M.}~\bibnamefont{Ruggieri}},
\bibnamefont{and}
\bibinfo{author}{\bibfnamefont{R.}~\bibnamefont{Gatto}},
  \bibinfo{journal}{Phys.\ Lett.\ B} \textbf{\bibinfo{volume}{657}},
  \bibinfo{pages}{64} (\bibinfo{year}{2007}).

\bibitem[{\citenamefont{Ratti et al.}(2007)}]{Ratti2}
\bibinfo{author}{\bibfnamefont{C.}~\bibnamefont{Ratti}},
\bibinfo{author}{\bibfnamefont{S.}~\bibnamefont{R\"{o}{\ss}ner}},
\bibinfo{author}{\bibfnamefont{M.}~\bibfnamefont{A.}~\bibnamefont{Thaler}},
\bibnamefont{and}
\bibinfo{author}{\bibfnamefont{W.}~\bibnamefont{Weise}},  
  \bibinfo{journal}{Eur. Phys. J.\ C} \textbf{\bibinfo{volume}{49}},
  \bibinfo{pages}{213} (\bibinfo{year}{2007}). 

\bibitem[{\citenamefont{Rossner et al.}(2007)}]{Rossner}
\bibinfo{author}{\bibfnamefont{S.}~\bibnamefont{R\"{o}{\ss}ner}},
\bibinfo{author}{\bibfnamefont{C.}~\bibnamefont{Ratti}},
\bibnamefont{and}
\bibinfo{author}{\bibfnamefont{W.}~\bibnamefont{Weise}},  
  \bibinfo{journal}{Phys. Rev.\ D} \textbf{\bibinfo{volume}{75}},
  \bibinfo{pages}{034007} (\bibinfo{year}{2007}). 

\bibitem[{\citenamefont{Hansen et al.}(2007)}]{Hansen}
\bibinfo{author}{\bibfnamefont{H.}~\bibnamefont{Hansen}}, 
\bibinfo{author}{\bibfnamefont{W.}~\bibfnamefont{M.}~\bibnamefont{Alberico}},
\bibinfo{author}{\bibfnamefont{A.}~\bibnamefont{Beraudo}}, 
\bibinfo{author}{\bibfnamefont{A.}~\bibnamefont{Molinari}},
\bibinfo{author}{\bibfnamefont{M.}~\bibnamefont{Nardi}},
\bibnamefont{and}
\bibinfo{author}{\bibfnamefont{C.}~\bibnamefont{Ratti}}, 
  \bibinfo{journal}{Phys. Rev.\ D} \textbf{\bibinfo{volume}{75}},
  \bibinfo{pages}{065004} (\bibinfo{year}{2007}). 

\bibitem[{\citenamefont{Sasaki et al.}(2007)}]{Sasaki1}
\bibinfo{author}{\bibfnamefont{C.}~\bibnamefont{Sasaki}},
\bibinfo{author}{\bibfnamefont{B.}~\bibnamefont{Friman}},
\bibnamefont{and}
\bibinfo{author}{\bibfnamefont{K.}~\bibnamefont{Redlich}}, 
\bibinfo{journal}{Phys. Rev.\ D} \textbf{\bibinfo{volume}{75}},
  \bibinfo{pages}{074013} (\bibinfo{year}{2007}). 

\bibitem[{\citenamefont{Schaefer}(2007)}]{Schaefer}
\bibinfo{author}{\bibfnamefont{B.}~\bibfnamefont{J.}~\bibnamefont{Schaefer}},
\bibinfo{author}{\bibfnamefont{J.}~\bibfnamefont{M.}~\bibnamefont{Pawlowski}},
\bibnamefont{and}
\bibinfo{author}{\bibfnamefont{J.}~\bibnamefont{Wambach}},  
  \bibinfo{journal}{Phys.\ Rev.\  D} \textbf{\bibinfo{volume}{76}},
  \bibinfo{pages}{074023} (\bibinfo{year}{2007}).

\bibitem[{\citenamefont{Fu}(2007)}]{Fu}
\bibinfo{author}{\bibfnamefont{W.}~\bibfnamefont{J.}~\bibnamefont{Fu}},
\bibinfo{author}{\bibfnamefont{Z.}~\bibnamefont{Zhang}},
\bibnamefont{and}
\bibinfo{author}{\bibfnamefont{Y.}~\bibfnamefont{X.}~\bibnamefont{Liu}},
  \bibinfo{journal}{Phys.\ Rev.\  D} \textbf{\bibinfo{volume}{77}},
  \bibinfo{pages}{014006} (\bibinfo{year}{2008}).

\bibitem[{\citenamefont{Kashiwa et al}(2008)}]{Kashiwa3}
\bibinfo{author}{\bibfnamefont{K.}~\bibnamefont{Kashiwa}}, 
\bibinfo{author}{\bibfnamefont{H.}~\bibnamefont{Kouno}}, 
\bibinfo{author}{\bibfnamefont{M.}~\bibnamefont{Matsuzaki}}, 
\bibnamefont{and}
\bibinfo{author}{\bibfnamefont{M.}~\bibnamefont{Yahiro}},
  \bibinfo{journal}{Phys.\ Lett.\ B} \textbf{\bibinfo{volume}{662}},
  \bibinfo{pages}{26} (\bibinfo{year}{2008}).

\bibitem[{\citenamefont{Costa et al}(2008)}]{Costa}
\bibinfo{author}{\bibfnamefont{P.}~\bibnamefont{Costa}}, 
\bibinfo{author}{\bibfnamefont{C.}~\bibfnamefont{A.}~\bibfnamefont{de}~\bibnamefont{Sousa}}, 
\bibinfo{author}{\bibfnamefont{M.}~\bibfnamefont{C.}~\bibnamefont{Ruivo}}, 
\bibnamefont{and}
\bibinfo{author}{\bibfnamefont{H.}~\bibnamefont{Hansen}},
 \bibinfo{howpublished}{arXiv:hep-ph/0801.3616} (\bibinfo{year}{2008}).

\bibitem[{\citenamefont{Fukushima}(2008)}]{Fukushima2}
\bibinfo{author}{\bibfnamefont{K.}~\bibnamefont{Fukushima}}, 
\bibinfo{howpublished}{arXiv:hep-ph/0803.3318} (\bibinfo{year}{2008}).

\bibitem[{\citenamefont{Kashiwa et al}(2008)}]{Kashiwa4}
\bibinfo{author}{\bibfnamefont{K.}~\bibnamefont{Kashiwa}}, 
\bibinfo{author}{\bibfnamefont{Y.}~\bibnamefont{Sakai}}, 
\bibinfo{author}{\bibfnamefont{H.}~\bibnamefont{Kouno}}, 
\bibinfo{author}{\bibfnamefont{M.}~\bibnamefont{Matsuzaki}}, 
\bibnamefont{and}
\bibinfo{author}{\bibfnamefont{M.}~\bibnamefont{Yahiro}},
  \bibinfo{howpublished}{arXiv:hep-ph/0804.3557} (\bibinfo{year}{2008})

\bibitem[{\citenamefont{Abuki}(2008)}]{Abuki}
\bibinfo{author}{\bibfnamefont{H.}~\bibnamefont{Abuki}}, 
\bibinfo{author}{\bibfnamefont{M.}~\bibnamefont{Ciminale}}, 
\bibinfo{author}{\bibfnamefont{R.}~\bibnamefont{Gatto}}, 
\bibinfo{author}{\bibfnamefont{N.}~\bibfnamefont{D.}~\bibnamefont{Ippolito}}, 
\bibinfo{author}{\bibfnamefont{G.}~\bibnamefont{Nardulli}}, 
\bibnamefont{and}
\bibinfo{author}{\bibfnamefont{M.}~\bibnamefont{Ruggieri}}, 
  \bibinfo{howpublished}{arXiv:hep-ph/0801.4254} (\bibinfo{year}{2008});
\bibinfo{author}{\bibfnamefont{H.}~\bibnamefont{Abuki}}, 
\bibinfo{author}{\bibfnamefont{M.}~\bibnamefont{Ciminale}}, 
\bibinfo{author}{\bibfnamefont{R.}~\bibnamefont{Gatto}}, 
\bibinfo{author}{\bibfnamefont{G.}~\bibnamefont{Nardulli}}, 
\bibnamefont{and}
\bibinfo{author}{\bibfnamefont{M.}~\bibnamefont{Ruggieri}}, 
  \bibinfo{howpublished}{arXiv:hep-ph/0802.2396} (\bibinfo{year}{2008});
\bibinfo{author}{\bibfnamefont{H.}~\bibnamefont{Abuki}}, 
\bibinfo{author}{\bibfnamefont{R.}~\bibnamefont{Anglani}}, 
\bibinfo{author}{\bibfnamefont{R.}~\bibnamefont{Gatto}}, 
\bibinfo{author}{\bibfnamefont{G.}~\bibnamefont{Nardulli}}, 
\bibnamefont{and}
\bibinfo{author}{\bibfnamefont{M.}~\bibnamefont{Ruggieri}}, 
  \bibinfo{howpublished}{arXiv:hep-ph/0805.1509} (\bibinfo{year}{2008});
\bibinfo{author}{\bibfnamefont{H.}~\bibnamefont{Abuki}}, 
 \bibinfo{howpublished}{arXiv:hep-ph/0805.3076} (\bibinfo{year}{2008}).

\bibitem[{\citenamefont{Asakawa and Yazaki}(1989)}]{AY}
\bibinfo{author}{\bibfnamefont{M.}~\bibnamefont{Asakawa}} \bibnamefont{and}
  \bibinfo{author}{\bibfnamefont{K.}~\bibnamefont{Yazaki}},
  \bibinfo{journal}{Nucl.\ Phys.} \textbf{\bibinfo{volume}{A504}},
  \bibinfo{pages}{668} (\bibinfo{year}{1989}). 

\bibitem[{\citenamefont{Kitazawa et al. }(2002)}]{KKKN}
\bibinfo{author}{\bibfnamefont{M.}~\bibnamefont{Kitazawa}},
{\bibfnamefont{T.}~\bibnamefont{Koide}},
{\bibfnamefont{T.}~\bibnamefont{Kunihiro}},
\bibnamefont{and}
\bibinfo{author}{\bibfnamefont{Y.}~\bibnamefont{Nemoto}},
\bibinfo{journal}{Prog. Theor. Phys.} \textbf{\bibinfo{volume}{108}},
\bibinfo{pages}{929} (\bibinfo{year}{2002}). 

\bibitem[{\citenamefont{Kashiwa et al}(2006)}]{Kashiwa}
\bibinfo{author}{\bibfnamefont{K.}~\bibnamefont{Kashiwa}}, 
\bibinfo{author}{\bibfnamefont{H.}~\bibnamefont{Kouno}}, 
\bibinfo{author}{\bibfnamefont{T.}~\bibnamefont{Sakaguchi}}, 
\bibinfo{author}{\bibfnamefont{M.}~\bibnamefont{Matsuzaki}}, 
\bibnamefont{and}
\bibinfo{author}{\bibfnamefont{M.}~\bibnamefont{Yahiro}},
\bibinfo{journal}{Phys. Lett.\ B} \textbf{\bibinfo{volume}{647}},
\bibinfo{pages}{446} (\bibinfo{year}{2007}); 
\bibinfo{author}{\bibfnamefont{K.}~\bibnamefont{Kashiwa}}, 
\bibinfo{author}{\bibfnamefont{M.}~\bibnamefont{Matsuzaki}}, 
\bibinfo{author}{\bibfnamefont{H.}~\bibnamefont{Kouno}}, 
\bibnamefont{and}
\bibinfo{author}{\bibfnamefont{M.}~\bibnamefont{Yahiro}},
\bibinfo{journal}{Phys. Lett.\ B} \textbf{\bibinfo{volume}{657}},
\bibinfo{pages}{143} (\bibinfo{year}{2007}). 

\bibitem[{\citenamefont{Walecka}(1974)}]{Walecka}
\bibinfo{author}{\bibfnamefont{J. D.}~\bibnamefont{Walecka}},
\bibinfo{journal}{Ann. Phys. } \textbf{\bibinfo{volume}{83}},
\bibinfo{pages}{491} (\bibinfo{year}{1974}).

\bibitem[{\citenamefont{Kashiwa and Sakaguchi}(2003)}]{KS}
\bibinfo{author}{\bibfnamefont{T.}~\bibnamefont{Kashiwa}} \bibnamefont{and}
\bibinfo{author}{\bibfnamefont{T.}~\bibnamefont{Sakaguchi}},
\bibinfo{journal}{Phys.\ Rev.\ D} \textbf{\bibinfo{volume}{68}},
\bibinfo{pages}{065002} (\bibinfo{year}{2003}).

\bibitem[{\citenamefont{Sakaguchi et al}(2006)}]{Sakaetal}
\bibinfo{author}{\bibfnamefont{T.}~\bibnamefont{Sakaguchi}}, 
\bibinfo{author}{\bibfnamefont{M.}~\bibnamefont{Matsuzaki}}, 
\bibinfo{author}{\bibfnamefont{H.}~\bibnamefont{Kouno}}, 
 \bibnamefont{and}
\bibinfo{author}{\bibfnamefont{M.}~\bibnamefont{Yahiro}},
\bibinfo{journal}{Centr.\ Eur.\ J.\ Phys.} \textbf{\bibinfo{volume}{6}},
\bibinfo{pages}{116} (\bibinfo{year}{2008}).

\bibitem[{\citenamefont{Kouno et al.}(2005)}]{KSKHTMY}
\bibinfo{author}{\bibfnamefont{H.}~\bibnamefont{Kouno}},
\bibinfo{author}{\bibfnamefont{T.}~\bibnamefont{Sakaguchi}},
\bibinfo{author}{\bibfnamefont{K.}~\bibnamefont{Kashiwa}},
\bibinfo{author}{\bibfnamefont{M.}~\bibnamefont{Hamada}},
\bibinfo{author}{\bibfnamefont{H.}~\bibnamefont{Tokudome}},
\bibinfo{author}{\bibfnamefont{M.}~\bibnamefont{Matsuzaki}},
\bibnamefont{and}
\bibinfo{author}{\bibfnamefont{M.}~\bibnamefont{Yahiro}},
\bibinfo{journal}{Soryushiron Kenkyu} \textbf{\bibinfo{volume}{112}}
\bibinfo{pages}{C67}
\bibinfo{howpublished}{(arXiv:nucl-th/0509057)} 
 (\bibinfo{year}{2005}).

\bibitem[{\citenamefont{Boyd et al.}(1996)}]{Boyd}
\bibinfo{author}{\bibfnamefont{G.}~\bibnamefont{Boyd}},
\bibinfo{author}{\bibfnamefont{J.}~\bibnamefont{Engels}},
\bibinfo{author}{\bibfnamefont{F.}~\bibnamefont{Karsch}},
\bibinfo{author}{\bibfnamefont{E.}~\bibnamefont{Laermann}},
\bibinfo{author}{\bibfnamefont{C.}~\bibnamefont{Legeland}},
\bibinfo{author}{\bibfnamefont{M.}~\bibnamefont{L\"{u}tgemeier}},
\bibnamefont{and}
\bibinfo{author}{\bibfnamefont{B.}~\bibnamefont{Petersson}},
 \bibinfo{journal}{Nucl. Phys.} \textbf{\bibinfo{volume}{B469}},
\bibinfo{pages}{419} (\bibinfo{year}{1996}). 

\bibitem[{\citenamefont{Kaczmarek}(2002)}]{Kaczmarek}
\bibinfo{author}{\bibfnamefont{O.}~\bibnamefont{Kaczmarek}},
\bibinfo{author}{\bibfnamefont{F.}~\bibnamefont{Karsch}},
\bibinfo{author}{\bibfnamefont{P.}~\bibnamefont{Petreczky}},
\bibnamefont{and}
\bibinfo{author}{\bibfnamefont{F.}~\bibnamefont{Zantow}},  
  \bibinfo{journal}{Phys. Lett.\ B} \textbf{\bibinfo{volume}{543}},
  \bibinfo{pages}{41} (\bibinfo{year}{2002}).

\bibitem[{\citenamefont{Karsch}(2002)}]{Karsch3}
\bibinfo{author}{\bibfnamefont{F.}~\bibnamefont{Karsch}}, 
\bibinfo{journal}{Lect. notes Phys. } \textbf{\bibinfo{volume}{583}},
\bibinfo{pages}{209} (\bibinfo{year}{2002}). 

\bibitem[{\citenamefont{Karsch, Leermann and Peikert}(2002)}]{Karsch4}
\bibinfo{author}{\bibfnamefont{F.}~\bibnamefont{Karsch}}, 
\bibinfo{author}{\bibfnamefont{E.}~\bibnamefont{Laermann}}, 
\bibnamefont{and}
\bibinfo{author}{\bibfnamefont{A.}~\bibnamefont{Peikert}},
\bibinfo{journal}{Nucl. Phys. \ B} \textbf{\bibinfo{volume}{605}},
\bibinfo{pages}{579} (\bibinfo{year}{2002}). 

\bibitem[{\citenamefont{Cheng et al.}(2006)}]{MCheng}
\bibinfo{author}{\bibfnamefont{M.}~\bibnamefont{Cheng et al.}}, 
\bibinfo{journal}{Phys. Rev. \ D} \textbf{\bibinfo{volume}{74}},
\bibinfo{pages}{054507} (\bibinfo{year}{2006}). 

\end{thebibliography}
\end{document}